\documentclass[a4paper,11pt]{article}
\usepackage{jcappub} 
\usepackage{aas_macros}
\usepackage{lineno}
\usepackage{booktabs}
\usepackage{graphicx}
\usepackage{amsthm,amsmath,amssymb}
\usepackage{mathrsfs}
\usepackage{hyperref}

\title{\boldmath Synergy between the gravitational potential decay rate and other structure growth probes in testing gravity}

\author[a,c,d]{Shang Li,}
\author[a,b,c,d]{Pengjie Zhang,} 
\author[e]{Fuyu Dong}
\affiliation[a]{Department of Astronomy, School of Physics and Astronomy, Shanghai Jiao Tong University, Shanghai, 200240, People’s Republic of China}
\affiliation[b]{Division of Astronomy and Astrophysics, Tsung-Dao Lee Institute, Shanghai Jiao Tong University, Shanghai, 200240, People’s Republic of China}
\affiliation[c]{State Key Laboratory of Dark Matter Physics, Shanghai 200240, People’s Republic of China}  
\affiliation[d]{Key Laboratory for Particle Astrophysics and Cosmology (MOE)/Shanghai Key Laboratory for Particle Physics and Cosmology, People’s Republic of China}
\affiliation[e]{South-Western Institute for Astronomy Research, Yunnan University, Kunming 650500, China}
\emailAdd{shanglicosmo@sjtu.edu.cn}
\emailAdd{zhangpj@sjtu.edu.cn}

\abstract{We test gravity by exploiting the synergy between the gravitational potential decay rate ($\mathit{DR}$) and complementary structure-growth probes: these observables respond to MG parameters with different degeneracy directions, so their combination yields stronger constraints than any single probe. We adopt the tomographic $\mathit{DR}$ measurements reported in \citep{2025ApJ...982...99D} and combine them with CMB lensing tomography $\Sigma_8$ measurements and $f\sigma_8$ measurements from DESI DR1 full-shape analyses and the DESI peculiar-velocity field. We apply this joint data vector to two representative frameworks: phenomenological parameterizations and the Effective Field Theory (EFT) $\alpha$-basis.
For the phenomenological form $P_{\rm MG}(a)=1+P_{{\rm MG},0}\,\Omega_{\rm DE}(a)/\Omega_{\rm DE}(0)$, where $P_{\rm MG}$ denotes $\mu$, $\eta$, or $\Sigma$, we obtain $\mu_0=0.09\pm0.35$ and $\Sigma_0=0.01\pm0.06$. Compared to the measurements combination $\Sigma_8+f\sigma_8$, including $\mathit{DR}$ tightens the constraint on $\Sigma_0$ by a factor of $\sim2$. For the $(\mu_0,\eta_0)$ case we find $\mu_0=0.06^{+0.17}_{-0.23}$ and $\eta_0=-0.03^{+0.36}_{-0.46}$; relative to $\Sigma_8+f\sigma_8$, adding $\mathit{DR}$ improves the constraints on both parameters by a factor of $\sim1.5$.
In the EFT $\alpha$-basis, adopting the parameterization $\alpha_i(a)=c_i\,\Omega_{\rm DE}(a)$ with $i\in\{{\rm M,B}\}$, we find $c_{\rm M}=0.64^{+0.32}_{-0.72}$ and $c_{\rm B}=0.31^{+0.19}_{-0.29}$. The corresponding EFT uncertainties are about a factor of $\sim2$ smaller than those reported in \citep{2025JCAP...09..053I}, which combined DESI full-shape and BAO measurements with DES-SN5YR and CMB data. These results demonstrate the capability of $\mathit{DR}$  and the necessity of including the $\mathit{DR}$ measurements in testing gravity. }

\begin{document}
\maketitle
\flushbottom

\section{Introduction}\label{sec:intro}
The physical origin of cosmic acceleration remains a central open question in modern cosmology \citep{1998AJ....116.1009R,1999ApJ...517..565P}. Although a cosmological constant within $\Lambda$CDM continues to provide an excellent fit to many data sets \citep{2020A&A...641A...6P,2021PhRvD.103h3533A}, its microscopic interpretation is unsettled, motivating both dynamical dark energy (DE) and modified gravity (MG) as viable explanations. Crucially, MG scenarios can alter not only the background expansion but also the evolution of linear perturbations, changing the growth of structure and the time dependence of metric potentials  \citep{2012PhR...513....1C, 2016RPPh...79d6902K, 2019LRR....22....1I, 2019ARA&A..57..335F}. A wide range of measurements, including Type~Ia supernovae (SN~Ia), baryon acoustic oscillations (BAO), redshift space distortions (RSD), weak lensing, and galaxy clustering, has been used to constrain DE/MG models (e.g., \citep{2022ApJ...938..110B,2025JCAP...02..021A,2025PhRvD.112h3515A,2025A&A...702A.169S,2026arXiv260114559D, 2026arXiv260210065D,2018MNRAS.480.3725S,2021PhRvD.103h3533A,2025JCAP...09..053I,2020JCAP...07..015N,2023PhRvD.107h3504A,2020A&A...642A.158B,2017MNRAS.471.1259J}). Among these observables, those that directly track the time evolution of the gravitational potential are particularly discriminating, because potential evolution is a generic and often distinctive imprint of late-time acceleration physics.

A direct probe of late-time potential evolution is provided by the integrated Sachs--Wolfe (ISW) effect \citep{1967ApJ...147...73S}, generated when CMB photons traverse time-varying gravitational potentials. In linear theory, the ISW temperature anisotropy depends on the time derivative of the Weyl potential, $(\dot{\Phi}+\dot{\Psi})$, and therefore encodes the decay rate of the gravitational potential, hereafter denoted as $\mathit{DR}$. This information is highly sensitive to gravity: in GR the potential is nearly constant during matter domination but evolves once acceleration becomes important \citep{2015PhR...568....1J,2016RPPh...79d6902K}. Observationally, ISW information is most commonly extracted via cross correlations between CMB temperature maps and large scale structure (LSS) tracers \citep{2016A&A...594A..21P,2021MNRAS.500.3838D,2022MNRAS.517.3785B,2022JCAP...09..002K,2024PhRvD.110l3525S,2025PhRvD.112h3537C}. However, the interpretation of the measured correlation signal is complicated by uncertainties in galaxy bias and the matter power spectrum, which can bias and/or degrade the resulting constraints on the nature of DE and gravity.

To mitigate these limitations, \cite{2006ApJ...647...55Z} proposed a strategy to isolate $\mathit{DR}$ by combining the ISW--galaxy cross-correlation $C_{\ell}^{Ig}$ with the lensing--galaxy cross-correlation $C_{\ell}^{\kappa g}$. The basic idea is to construct a $\mathit{DR}$ estimator that depends on a ratio of the two cross-spectra,
\begin{equation}
C_{\ell}^{Ig}\simeq \mathit{DR} (z_m)C_{\ell}^{\kappa g},
\label{eq:DR_estimator}
\end{equation}
where $z_m$ is the mean redshift of a chosen galaxy redshift bin. This construction has two notable advantages. First, the resulting statistic depends on a minimal set of cosmological quantities: beyond the DE equation-of-state parameter(s) or MG parameters of interest, it involves primarily the matter density parameter $\Omega_{\rm m}$, substantially reducing degeneracies with the broader cosmological parameter space. Second, $\mathit{DR}$ can be more sensitive to DE/MG physics than many conventional late-time observables; for example, \cite{2022ApJ...938...72D} found that its sensitivity to the DE equation-of-state parameter $w$ is approximately a factor of $3/4/5$ higher at $z=0.3/0.5/0.7$ than that of the Hubble parameter $H(z)$. These features make $\mathit{DR}$ a promising and comparatively clean avenue for testing gravity with current and upcoming surveys.

Complementary to potential evolution probes, measurements of the growth of structure provide an independent and highly informative test of gravity. MG generically alters the relation between the matter density field and the potentials, leading to changes in the linear growth factor and growth rate that can be captured by observables such as $\sigma_8$ and $f\sigma_8$. Importantly, while $\mathit{DR}$ directly targets potential time variation, it cannot break degeneracies in multi-parameter MG models. Growth observables provide independent sensitivity to the matter fluctuations and can therefore help break these degeneracies. This motivates a joint analysis that combines $DR$ with structure growth measurements to obtain more informative constraints on parameterized MG frameworks.

In this work, we provide joint constraints on parameterized MG frameworks by combining the recent tomographic $\mathit{DR}$ measurement with complementary structure growth information. We consider two phenomenological descriptions as well as the  Effective Field Theory (EFT) constraints in the $\alpha$-basis. Throughout this work, we assume a spatially flat $\Lambda$CDM background expansion and restrict the MG modifications to the perturbation sector. The paper outline is as follows. Sect.~\ref{sec:formalism} introduces the parameterized MG models considered in our analysis. In Sect.~\ref{sec:data} we describe the datasets used in this work and the motivation for our data selection. Our constraints are presented in Sect.~\ref{sec:results}, and we summarize our findings in Sect.~\ref{sec:summary}. Appendix~\ref{app:contours} contains some supplementary results.

\section{Formalism}\label{sec:formalism}
\subsection{Phenomenological parameterization}
\label{sec:mg_param}

We introduce phenomenological MG functions directly at the level of the linearly perturbed Einstein field equations. In the conformal Newtonian gauge, the flat Friedmann--Lema\^{\i}tre--Robertson--Walker(FLRW) metric with scalar perturbations is written as
\begin{equation}
ds^{2} = a^{2}(\tau)\Big[-(1+2\Psi)\,d\tau^{2} + (1-2\Phi)\,\delta_{ij}\,dx^{i}dx^{j}\Big],
\label{eq:newtonian_gauge_metric}
\end{equation}
where $\Phi$ and $\Psi$ are the two gravitational potentials and $\tau$ is conformal time.

In Fourier space, and in the late-time Universe where anisotropic stresses are negligible, the relativistic Poisson equation can be written in the MG form
\begin{equation}
k^{2}\Psi = -4\pi G a^{2}\,\mu(a)\sum_{i}\rho_{i}\Delta_{i},
\label{eq:poisson_mu}
\end{equation}
where $\rho_i$ is the density of species $i$ and $\Delta_i$ is the gauge-invariant rest-frame overdensity. The function $\mu(a)$ encodes an effective modification of the gravitational coupling felt by massive particles and therefore governs the growth of linear structure.

In GR, $\Phi\simeq\Psi$ at late times (neglecting anisotropic stress), whereas in MG they can differ. This departure is commonly characterized by the gravitational slip parameter
\begin{equation}
\eta(a) \equiv \frac{\Phi}{\Psi}.
\label{eq:slip_eta}
\end{equation}

Finally, the motion of massless particles (and hence gravitational lensing observables) is governed by the Weyl potential $\Phi+\Psi$. Combining the perturbed field equations yields
\begin{equation}
k^{2}\left(\Phi+\Psi\right) = -8\pi G a^{2}\,\Sigma(a)\sum_{i}\rho_{i}\Delta_{i},
\label{eq:weyl_sigma}
\end{equation}
\setcounter{footnote}{0}
where $\Sigma(a)$ parameterize modifications to the lensing potential relative to GR. Note that $\mu, \Sigma, and\ \eta$ are not independent. At low redshift, these MG parameters satisfy\footnote{In general, $\mu$, $\Sigma$, and $\eta$ may depend on both time(a) and scale(k); here we consider time dependence only.}
\begin{equation}
\Sigma(a) = \frac{\mu(a)}{2}\left[\eta(a)+1\right].
\label{eq:sigma_mu_eta_relation}
\end{equation}
In GR, one has $\mu=\Sigma=\eta=1$, recovering the standard perturbed Einstein equations.

We emphasize that the $\mu$, $\Sigma$, and $\eta$ description is defined at the level of linear perturbations, and is therefore most robustly applied on (quasi-)linear scales with appropriate scale cuts.

Building on the above parameterization, the $\mathit{DR}$ observable can be written as
\begin{equation}
\mathit{DR}(z)=\left(1-f-\frac{d\ln\Sigma}{d\ln a}\right)\left(\frac{a\,H(z)/c}{W_L(z)}\right),
\label{eq:DR_muSigma}
\end{equation}
where $f\equiv d\ln D/d\ln a$ is the linear growth rate and $D(a)$ is the linear growth factor. $W_L(z)=[1-\chi(z)/\chi_\ast]/\chi(z)$ is the CMB lensing kernel, where $\chi(z)$ and $\chi_\ast$ denote the comoving distances to the lens redshift and the last-scattering surface, respectively. We obtain $D(a)$ numerically by solving
\begin{equation}
D'' + D'\left(\frac{H'}{H}+\frac{3}{a}\right)
-\frac{3}{2}\frac{\Omega_m(a)}{a^2}\,D\,[1+\mu(a)] = 0,
\label{eq:growth_mu}
\end{equation}
where primes denote derivatives with respect to $a$.

For specific parameterizations, we note that cosmic acceleration is a late-time phenomenon; therefore, as $a\rightarrow 0$ the modified-gravity functions should approach their GR values, i.e., $P_{\rm MG}\rightarrow 1$, where $P_{\rm MG}$ denotes $\mu$, $\eta$, or $\Sigma$. The time dependence is often parameterized by assuming proportionality to the evolution of the dark-energy density parameter $\Omega_{\rm DE}(a)$. We therefore adopt
\begin{equation}
P_{\rm MG}(a)=1+P_{{\rm MG},0}\,\frac{\Omega_{\rm DE}(a)}{\Omega_{\rm DE}(0)}\,,
\label{eq:PMG_ode}
\end{equation}
where $P_{\rm MG,0}$ denotes $\mu_0$, $\eta_0$, or $\Sigma_0$. In GR, $P_{{\rm MG},0}=0$.

\subsection{EFT parameterization (\texorpdfstring{$\alpha$-basis}{alpha-basis})}\label{sec:alpha_basis}

Beyond the phenomenological description, we also consider the Effective Field Theory (EFT) of dark energy\cite{2015LNP...892...97T,2020PhR...857....1F}, which offers a flexible and comprehensive language to capture a broad range of MG models at the level of linear perturbations. In particular, we work in the $\alpha$-basis, a systematic framework associated with the Horndeski class of scalar--tensor theories.

In this basis, departures from GR in the linear dynamics are described by four time-dependent functions,
$\{\alpha_{\rm M}(t),\alpha_{\rm B}(t),\alpha_{\rm K}(t),\alpha_{\rm T}(t)\}$.
Here $\alpha_{\rm M}$ quantifies the running of the effective Planck mass,
\begin{equation}
\alpha_{\rm M}\equiv \frac{d\ln M_*^2}{d\ln a},
\end{equation}
where $M_*^2$ denotes the effective Planck masses. A running Planck mass modifies the growth of structures, introduces anisotropic stress and modifies the friction term in the gravitational wave equation.
$\alpha_{\rm B}$ describes the kinetic mixing (``braiding'') between the scalar and metric sectors. It is different from zero for all the theories showing non-minimal coupling to gravity.
$\alpha_{\rm K}$ controls the scalar kinetic contribution (``kineticity''), and it affects the speed of propagation of the DE field and is the only coupling present in quintessence or DE models.
$\alpha_{\rm T}\equiv c_T^2-1$ is the tensor speed excess and describes the deviation of the speed of propagation of gravitational waves from the speed of light. It affects the evolution of the scalar gravitational potentials leading to anisotropic stress.
To model their late-time evolution, we adopt the commonly used ansatz
\begin{equation}
\alpha_i(a)=c_i\,\Omega_{\rm DE}(a), \qquad i\in\{{\rm M,B,K,T}\},
\label{eq:alpha_param}
\end{equation}
which ensures $\alpha_i\rightarrow 0$ at early times as $\Omega_{\rm DE}(a)\rightarrow 0$. In this work, we fix $\alpha_{\rm T}=0$ at all times motivated by the stringent bound on the gravitational-wave speed from GW170817 and its electromagnetic ($\gamma$-ray) counterpart GRB170817A\cite{2017ApJ...848L..13A}. We also fix $c_{\rm K}=10^{-2}$ since current data typically have limited sensitivity to this function\cite{2016JCAP...06E.001B,2019MNRAS.482.3274R}.

On sub-horizon linear scales, and under the quasi-static approximation (QSA), the $\alpha$-functions can be mapped onto the phenomenological functions $\mu(z,k)$ and $\Sigma(z,k)$ introduced above.
\begin{align}
\mu(z) &=
\frac{M_{\mathrm{Pl}}^2}{M_*^2}\left[
1+\frac{2\left(\alpha_{\mathrm{M}}+\tfrac{1}{2}\alpha_{\mathrm{B}}\right)^2}
{c_s^2\left(\alpha_{\mathrm{K}}+\tfrac{3}{2}\alpha_{\mathrm{B}}^2\right)}
\right],
\label{eq:mu_alpha_qsa}
\\
\Sigma(z) &=
\frac{M_{\mathrm{Pl}}^2}{M_*^2}\left[
1+\frac{\left(\alpha_{\mathrm{M}}+\tfrac{1}{2}\alpha_{\mathrm{B}}\right)\left(\alpha_{\mathrm{M}}+\alpha_{\mathrm{B}}\right)}
{c_s^2\left(\alpha_{\mathrm{K}}+\tfrac{3}{2}\alpha_{\mathrm{B}}^2\right)}
\right].
\label{eq:sigma_alpha_qsa}
\end{align}
where $M_{\rm Pl}^2$ denotes the bare Planck masses. We also enforce the standard stability requirements to ensure a viable theory: the no-ghost condition demands $\alpha_{\rm K}+\frac{3}{2}\alpha_{\rm B}^2>0$,
while the absence of gradient instabilities requires a positive scalar sound speed, $c_s^2>0$, with $c_s^2$ computed following\cite{2019PhRvD..99j3502N,2021JCAP...01..013B,2021JCAP...09..024B}
\begin{equation}
\label{eq:cs2_def}
c_s^2 =
\frac{1}{\alpha_K+\frac{3}{2}\alpha_B^2}
\Bigg[
(2-\alpha_B)\left(\frac{1}{2}\alpha_B+\alpha_M-\frac{\dot{H}}{aH^2}\right)
-\frac{3(\rho_{\rm tot}+p_{\rm tot})}{H^2M_*^2}
+\frac{\dot{\alpha}_B}{aH}
\Bigg] \, .
\end{equation}

Following \cite{2025JCAP...09..053I}, we also consider three representative subclasses: (i) a maximally flexible case where both $\alpha_{\rm M}$ and $\alpha_{\rm B}$ vary according to Eq.~\eqref{eq:alpha_param}; (ii) a ``no-braiding'' subclass with $\alpha_{\rm B}=0$; and (iii) a ``no-slip'' subclass satisfying $\alpha_{\rm B}=-2\alpha_{\rm M}$ (corresponding to $\Phi=\Psi$), for which $\Sigma=\mu=M_{Pl}^2/M_*^2$. Viability conditions, including ghost and gradient stability, are imposed throughout the parameter inference($\alpha_K+\frac{3}{2}\alpha_B^2>0$ and $c_s^2>0$).

\section{Data}\label{sec:data}
In this section, we describe the datasets used in our analysis. We adopt a joint data vector consisting of the $\mathit{DR}$ measurement together with complementary structure-growth observables. While $\mathit{DR}$ provides a comparatively clean probe of late-time potential evolution, the growth data supply additional information needed to mitigate parameter degeneracies in multi-parameter MG models. All datasets used in this work are listed in Table~\ref{tab:data_all}.
\begin{figure}[t]
\centering
\includegraphics[width=\textwidth]{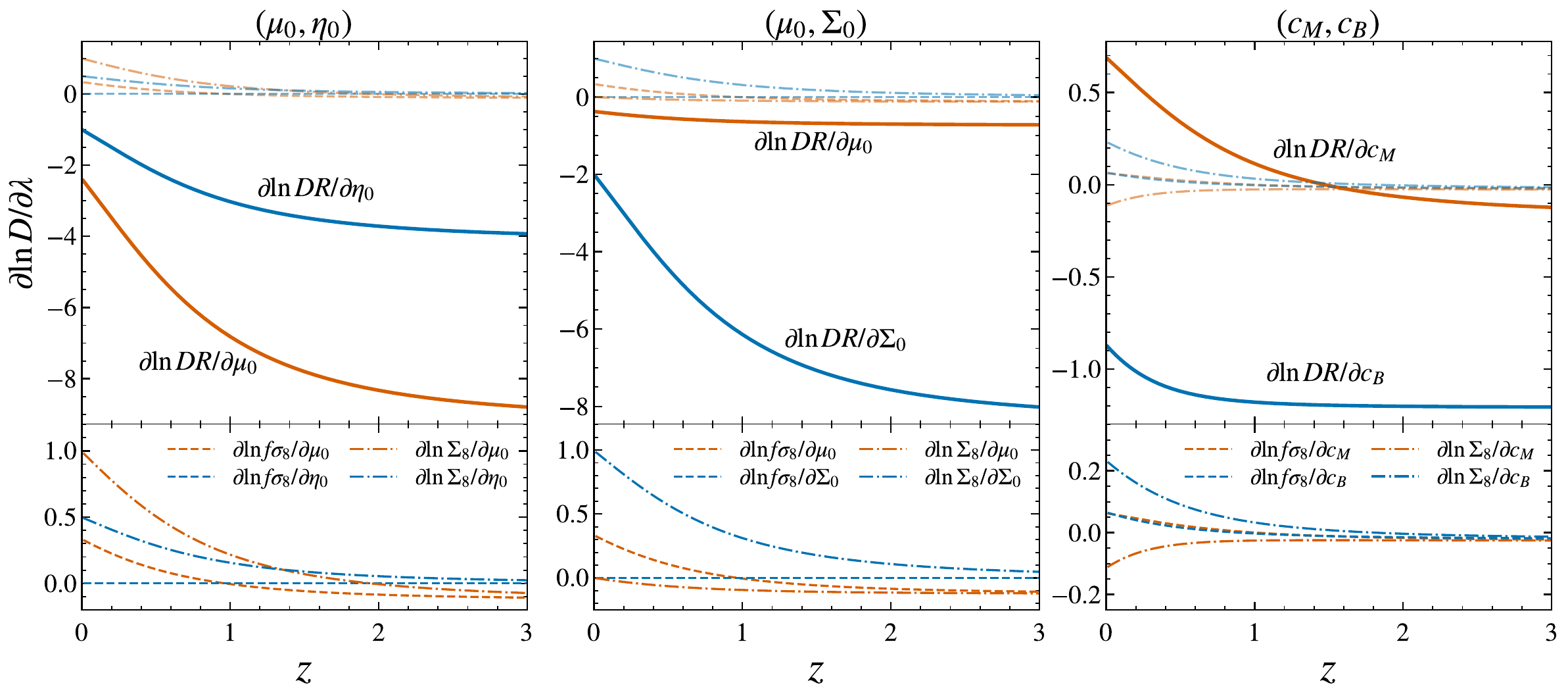}
\caption{Redshift-dependent sensitivities of the observables to MG parameters, quantified by $\partial\ln D/\partial\lambda$ and evaluated around the fiducial $\Lambda$CDM model, where $D\in\{\mathit{DR},\Sigma_8,f\sigma_8\}$. Columns correspond to the $(\mu_0,\eta_0)$, $(\mu_0,\Sigma_0)$, and $(c_{\rm M},c_{\rm B})$ parameterizations (left to right). In each column, the upper panel shows the sensitivities of all observables, while the lower panel provides a zoomed view highlighting the $\Sigma_8$ and $f\sigma_8$ sensitivities (note the different y-axis scale). $\mathit{DR}$ exhibits substantially larger sensitivities than $\Sigma_8$ and $f\sigma_8$, underscoring its strong constraining power. For a given probe, the relative signs of the sensitivities indicate the local degeneracy direction at fixed observable: same-sign responses correspond to an anti-correlation, whereas opposite-sign responses correspond to a positive correlation. Differences in these sign patterns across probes therefore imply partially orthogonal degeneracy directions, illustrating the complementarity (synergy) between $\mathit{DR}$ and growth measurements in constraining MG parameters.}
\label{fig:sensitivity}
\end{figure}

\subsection{\texorpdfstring{$\mathit{DR}$ }{DR}}\label{subsec:data_dr}

As introduced in Sect.~\ref{sec:intro}, $\mathit{DR}$  can be isolated by taking an appropriate ratio between the ISW--galaxy cross-correlation and the lensing--galaxy cross-correlation. Physically, the ISW contribution to the CMB temperature anisotropy is sourced by the time derivative of the (Weyl) potential, such that on the relevant large scales
\begin{equation}
\Delta T_{\rm ISW}\propto \dot{\Phi}+\dot{\Psi},
\end{equation}
where the dot denotes a derivative with respect to conformal time. In contrast, weak gravitational lensing is sourced by the potential itself, so that the lensing convergence satisfies schematically $\kappa \propto \nabla^2(\Phi+\Psi)$. Combining these two ingredients, the $\mathit{DR}$ observable can be expressed as a dimensionless decay rate of the Weyl potential \cite{2006ApJ...647...55Z}:
\begin{equation}
\mathit{DR}(z)=\left(-\frac{{\rm d}\ln D_{\Phi+\Psi}}{{\rm d}\ln a}\right)\left(\frac{aH(z)/c}{W_L(z)}\right),
\label{eq:DR_def}
\end{equation}
where $W_L(z)=[1-\chi(z)/\chi(z_s)]/\chi(z)$ is the lensing kernel, and $\chi(z)$ is the comoving radial distance to redshift $z$. Here $z_s$ is the source redshift (for CMB lensing, $z_s \simeq 1100$), and $D_{\Phi+\Psi}$ is the linear growth factor of the Weyl potential.

Figure~\ref{fig:sensitivity} summarizes the redshift-dependent sensitivities of $\mathit{DR}$ to the MG parameters considered in this work, including the phenomenological parameterizations and the EFT $\alpha$-basis. For the $(\mu_0,\eta_0)$ and $(\mu_0,\Sigma_0)$ models, the sensitivities of $\mathit{DR}$ to the MG parameters are negative over the redshift range of interest. Notably, in the $(\mu_0,\Sigma_0)$ case the sensitivity of $\mathit{DR}$ to $\mu_0$ varies only weakly with redshift, and its redshift dependence is much smaller than that in the $(\mu_0,\eta_0)$ case.

This difference can be understood from the mapping between the two phenomenological parameterizations. In the $(\mu_0,\eta_0)$ framework, $\Sigma$ is not an independent function but is related to $\mu$ and $\eta$ through Eq.~\eqref{eq:sigma_mu_eta_relation}, so that $\Sigma(a)=\Sigma(\mu(a),\eta(a))$.\footnote{Hereafter, unless stated otherwise, we use $\mu$, $\eta$, and $\Sigma$ to denote the corresponding time-dependent functions in the $(\mu_0,\eta_0)$ or $(\mu_0,\Sigma_0)$ parameterizations, i.e., $\mu(a)=1+\mu_0T(a)$, $\eta(a)=1+\eta_0T(a)$, and $\Sigma(a)=1+\Sigma_0T(a)$, with $T(a)$ specified in the text.} Therefore, in the $(\mu_0,\eta_0)$ parameterization, $\mu_0$ affects $\mathit{DR}$ through two channels: it changes the growth rate $f$ via the modified growth equation, and it also enters the term ${\rm d}\ln\Sigma/{\rm d}\ln a$ through the derived Weyl-potential modification $\Sigma(\mu,\eta)$. By contrast, in the $(\mu_0,\Sigma_0)$ framework, $\mu_0$ mainly affects $\mathit{DR}$ through $f$, while $\Sigma_0$ enters through ${\rm d}\ln\Sigma/{\rm d}\ln a$.

Using Eq.~\eqref{eq:sigma_mu_eta_relation}, the derived Weyl-potential modification in the $(\mu_0,\eta_0)$ framework can be written as
\begin{equation}
\Sigma(a)=\frac{\mu(a)\,[1+\eta(a)]}{2}
=\left[1+\mu_0T(a)\right]\left[1+\frac{\eta_0}{2}T(a)\right],
\end{equation}
where $T(a)=\Omega_{\rm DE}(a)/\Omega_{\rm DE}(0)$ denotes the adopted time dependence. Since $\mathit{DR}$ depends on $\Sigma$ through ${\rm d}\ln\Sigma/{\rm d}\ln a$, the product form above implies that the $\mu_0$ and $\eta_0$ contributions enter additively in $\ln\Sigma$. Consequently, the $\mathit{DR}$ sensitivity to $\mu_0$ in the $(\mu_0,\eta_0)$ framework is expected to be approximately the sum of the sensitivities to $\mu_0$ and $\Sigma_0$ in the $(\mu_0,\Sigma_0)$ framework, while the sensitivity to $\eta_0$ is approximately half of the $\Sigma_0$ sensitivity. This trend is also evident from the sensitivity curves shown in Figure~\ref{fig:sensitivity}. This explains why the $\mathit{DR}$ sensitivity to $\mu_0$ is enhanced and shows a stronger redshift dependence in the $(\mu_0,\eta_0)$ framework.

In the EFT $\alpha$-basis, unlike the two phenomenological cases, the sensitivities of $\mathit{DR}$ to the two parameters have opposite signs. This implies that the $\mathit{DR}$ response to $(c_{\rm M},c_{\rm B})$ changes with redshift, so different tomographic bins favor different degeneracy directions. Consequently, combining $\mathit{DR}$ measurements across redshift can help reduce parameter degeneracies and yield tighter joint constraints on $(c_{\rm M},c_{\rm B})$. At the same time, $\mathit{DR}$ shows larger absolute sensitivities than $\Sigma_8$ and $f\sigma_8$, suggesting that it provides the strongest constraints on the EFT parameters in our analysis. By comparison, $f\sigma_8$ has the smallest sensitivities and is therefore expected to yield the weakest constraints when used alone.

Observationally, $\mathit{DR}$ was first measured by \cite{2022ApJ...938...72D} using DESI Legacy Imaging Surveys DR8 \cite{2019ApJS..242....8Z} together with \textit{Planck} data over $0.2<z<0.8$, and \cite{2025ApJ...982...99D} later extended the analysis to $z<1.4$ using DESI Legacy Imaging Surveys DR9 \cite{2021MNRAS.501.3309Z}. We adopt this DR9 tomographic dataset in our analysis: it provides six redshift bins spanning $0.2\le z<1.4$, and yields an overall detection significance of $3.1\sigma$. To control key systematics in the DR-related correlation-function measurements, \cite{2025ApJ...982...99D} explicitly accounted for imaging systematics in the galaxy sample (mitigated via machine-learning-based imaging weights; see, e.g., \cite{2023MNRAS.520..161X}) and lensing magnification bias in the measured cross-correlations. The corresponding $\mathit{DR}$ data points used in this work are shown in Figure~\ref{fig:data}.

\subsection{\texorpdfstring{$\Sigma_{8}(z)$}{S8z}}\label{subsubsec:data_sig}
In this section, $\Sigma_8(z)$ denotes the \emph{lensing-inferred} $\sigma_8$-type growth-amplitude measurement. We adopt the notation $\Sigma_8$ (rather than $\sigma_8$) to emphasize that lensing-based inferences depend on the Weyl potential and can therefore be modified in MG scenarios, and to avoid confusion with the standard matter-fluctuation amplitude defined in GR-based analyses. In our phenomenological framework, this quantity is modeled as
\begin{equation}
\Sigma_8(z)=\sigma_8(0)\,D(z)\,\Sigma(z),
\end{equation}
where $D(z)$ is the linear growth factor and $\Sigma(z)$ parameterizes modifications to the Weyl potential.

In the context of MG, $\sigma_{8}$ plays a central role as it quantifies the amplitude of matter fluctuations on $8\,h^{-1}\,\mathrm{Mpc}$ scales and is therefore directly tied to the growth of cosmic structure, which is generically altered when gravity is modified.

Measurements of the growth amplitude can be obtained through multiple approaches, including galaxy--CMB lensing analyses\cite{2025arXiv251009563R,2025JCAP...06..008S,2025PhRvD.111j3540S,2025JCAP...10..077D,2024JCAP...06..012P}, cosmic shear based analyses\cite{2023PhRvD.107h3504A,2023OJAp....6E..36D,2023MNRAS.520.5016L}, and galaxy cluster number counts\cite{2019MNRAS.489..401Z,2024PhRvD.110h3510B,2024PhRvD.110h3511S,2025A&A...696A...5A}. In galaxy--CMB lensing studies, often referred to as CMB lensing tomography, the relevant information is typically extracted from a joint analysis of galaxy clustering and galaxy--CMB lensing correlations, while shear based analyses combine galaxy clustering, galaxy--galaxy lensing, and cosmic shear, often named as $3\times2$pt analyses. In addition to the two approaches discussed above, cluster number counts have emerged as a widely used probe in recent years, constraining the growth amplitude through the abundance and redshift evolution of massive halos.

In this work, we exclusively adopt datasets based on the CMB lensing tomography approach. Compared with shear-based analyses, this approach avoids several source galaxy related systematics. In cosmic shear and galaxy-galaxy lensing measurements, the lensing kernels depend sensitively on the redshift distributions of the source galaxies, which are usually calibrated from photometric redshifts and can therefore introduce non-negligible uncertainties. By contrast, CMB lensing has a well-defined source plane at the surface of last scattering, ($z\simeq1100$). Since the usable source galaxies in current shear surveys are mostly limited to redshifts of order unity, the corresponding foreground lens samples must generally lie at lower redshifts. CMB lensing, with its much higher source redshift, can instead be combined with galaxy tracers over a substantially broader redshift range, including tracers beyond the typical reach of galaxy shear tomography. In addition, CMB lensing is largely free from intrinsic alignment systematics, because it probes the lensing-induced distortion of the CMB photon field rather than the intrinsic shapes of source galaxies. These features make CMB lensing tomography a clean and complementary probe of the redshift evolution of structure growth.

Figure~\ref{fig:sensitivity} also presents the sensitivities of $\Sigma_8$ to the parameters in the different MG parameterizations considered in this work. In both the $(\mu_0,\Sigma_0)$ and $(\mu_0,\eta_0)$ models, $\Sigma_8$ exhibits relatively weak sensitivity to the MG parameters over the redshift range of interest. This indicates that, by itself, $\Sigma_8$ provides only limited constraining power on these parameters, and its main role in our analysis is to provide complementary growth information when combined with $\mathit{DR}$ and $f\sigma_8$.

\begin{figure}[!t]
\centering
\includegraphics[width=\linewidth]{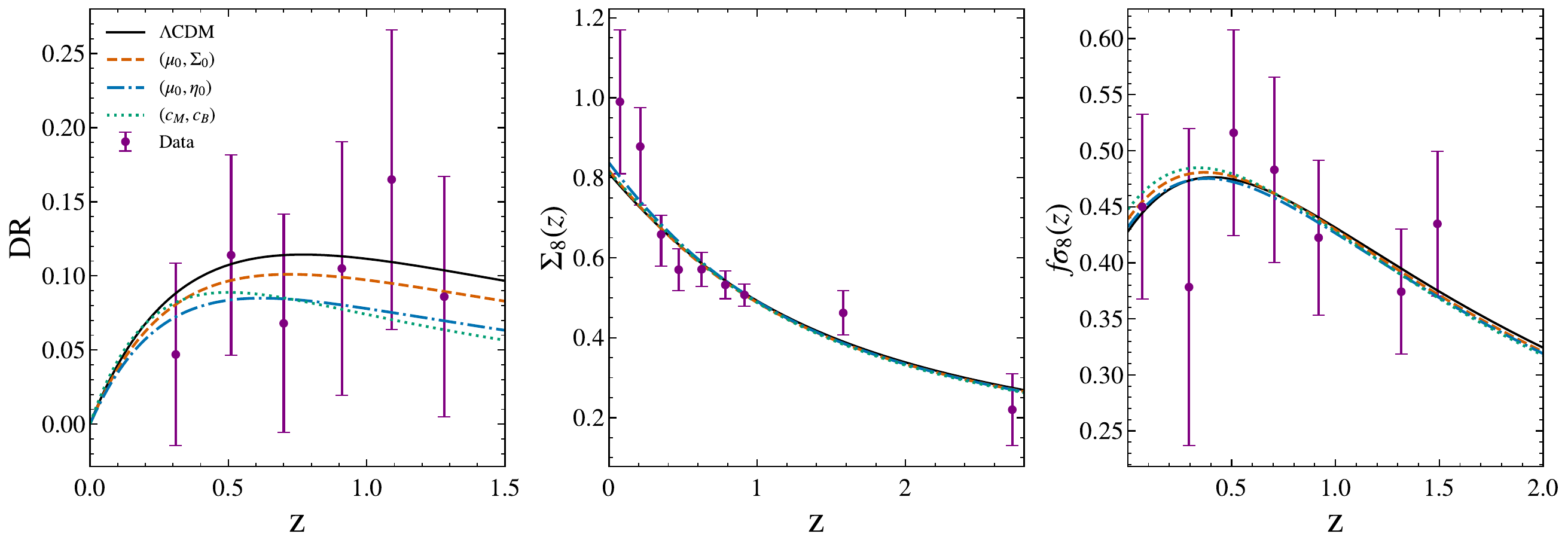}

\caption{The datasets used in this work, together with the best-fit predictions from our baseline joint analysis and the $\Lambda$CDM theoretical curves. Purple points with error bars show the measurements, the black solid line denotes the $\Lambda$CDM prediction, and the colored dashed lines show the best-fit predictions for the three MG parameterizations. For $\Sigma_8$ and $f\sigma_8$, the error bars shown have been inflated by $50\%$ to mitigate the potential model dependence. From left to right, the panels display DR, $\Sigma_8$, and $f\sigma_8$ as a function of redshift.}
\label{fig:data}
\end{figure}

Based on the above considerations, we adopt the following $\Sigma_8$ datasets in our analysis. We use the 2MPZ--\textit{Planck} cross-correlation to provide the lowest-redshift point ($z_{\rm eff}\simeq0.074$) \cite{2025arXiv251009563R}. We use DESI DR9 BGS and LRG samples cross-correlated with \textit{Planck}+\textit{ACT} to populate the intermediate-redshift range, with $z_{\rm eff}=\{0.211,\,0.352,\,0.470,\,0.625,\,0.785,\,0.914\}$ \cite{2025JCAP...06..008S,2025PhRvD.111j3540S}. Finally, we include two high-redshift tracer measurements: DESI DR1 QSOs cross-correlated with \textit{Planck} ($z_{\rm eff}\simeq1.6$) \cite{2025JCAP...10..077D} and Quaia cross-correlated with \textit{Planck} ($z_{\rm eff}\simeq2.7$) \cite{2024JCAP...06..012P}. Figure~\ref{fig:data} also compares the selected data points with the theoretical prediction under the $\Lambda$CDM model.
\subsection{\texorpdfstring{$f\sigma_{8}(z)$}{fs8}}
In addition to $\Sigma_8$, we also include measurements on $f\sigma_8$. Unlike the lensing-based $\Sigma_8$ measurements, $f\sigma_8$ is typically inferred from velocity field information. Consequently, within the $(\mu_0,\eta_0)$ and $(\mu_0,\Sigma_0)$ MG frameworks considered here, $f\sigma_8$ depends only on $\mu_0$ and can naturally help break parameter degeneracies. We also show the sensitivities of $f\sigma_8$ to the MG parameters in Fig.~\ref{fig:sensitivity}.

Measurements on $f\sigma_{8}$ can be obtained using a variety of cosmological probes, including peculiar velocities and RSD. Over the past years, numerous measurements of $f\sigma_{8}$ have been accumulated\cite{2017MNRAS.471.3135H, 2019MNRAS.487.5235Q,2020MNRAS.497.1275S,2021PhRvD.103h3533A,2023ApJ...958..180A,2024MNRAS.531...84B}; however, it is important to note that substantial overlap exists among many of these datasets, implying non-negligible covariance and preventing them from being treated as independent measurements. Such covariances are generally difficult to quantify. To avoid this issue, we only adopt the DESI DR1 full-shape results\cite{2025JCAP...09..008A,2025JCAP...07..028A} and the recently released constraints from the DESI peculiar-velocity field\cite{2025arXiv251203231Q,2025arXiv251203229L,2025arXiv251203230T}. We note that the DESI DR1 full-shape analyses report the ShapeFit observable $f\sigma_{s8}$ rather than the conventional $f\sigma_8$. In this work, we convert the published $f\sigma_{s8}$ measurements to $f\sigma_8$ following the prescription adopted in the DESI ShapeFit analyses. All $f\sigma_8$ values quoted and used in our likelihood therefore refer to the converted conventional quantity. The corresponding data are shown in Fig.~\ref{fig:data}.
\begin{table*}[t]
\centering
\caption{Measurements used in this work. Left: DR; middle: $f\sigma_8$; right: $\Sigma_8$.}
\label{tab:data_all}
\setlength{\tabcolsep}{8pt}
\renewcommand{\arraystretch}{1.15}
\begin{tabular}{cc cc cc}
\toprule
$z_{\rm eff}$ & $\mathit{DR}$ & $z_{\rm eff}$ & $f\sigma_8$ & $z_{\rm eff}$ & $\Sigma_8$ \\
\midrule
0.31 & $0.047^{+0.066}_{-0.057}$ & 0.070 & $0.450\pm0.055$ & 0.074 & $0.099\pm0.12$ \\
0.51 & $0.114^{+0.072}_{-0.063}$ & 0.295 & $0.378\pm0.094$ & 0.211 & $0.878^{+0.065}_{-0.098}$ \\
0.70 & $0.068^{+0.087}_{-0.060}$ & 0.510 & $0.516\pm0.061$ & 0.352 & $0.658^{+0.032}_{-0.053}$ \\
0.91 & $0.105^{+0.087}_{-0.084}$ & 0.706 & $0.483\pm0.055$ & 0.470 & $0.570\pm0.035$ \\
1.09 & $0.165^{+0.099}_{-0.103}$ & 0.919 & $0.422\pm0.046$ & 0.625 & $0.571\pm0.029$ \\
1.28 & $0.086^{+0.081}_{-0.081}$ & 1.317 & $0.374\pm0.037$ & 0.785 & $0.532\pm0.023$ \\
--   & --                          & 1.491 & $0.435\pm0.043$ & 0.914 & $0.507\pm0.019$ \\
--   & --                          & --    & --              & 1.58  & $0.462\pm0.037$ \\
--   & --                          & --    & --              & 2.72  & $0.220\pm0.060$ \\
\bottomrule
\end{tabular}
\end{table*}

\section{Results}\label{sec:results}
\begin{figure*}[t]
\centering
\includegraphics[width=0.3\textwidth]{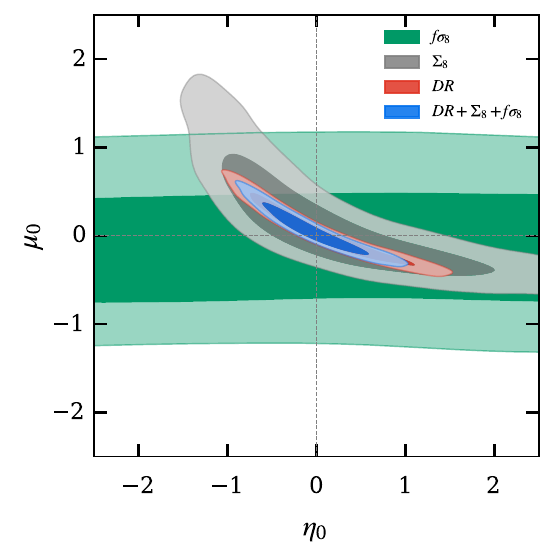}\hfill
\includegraphics[width=0.3\textwidth]{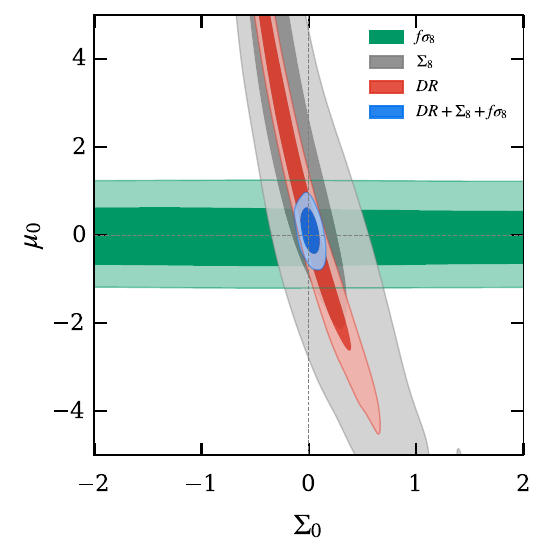}\hfill
\includegraphics[width=0.31\textwidth]{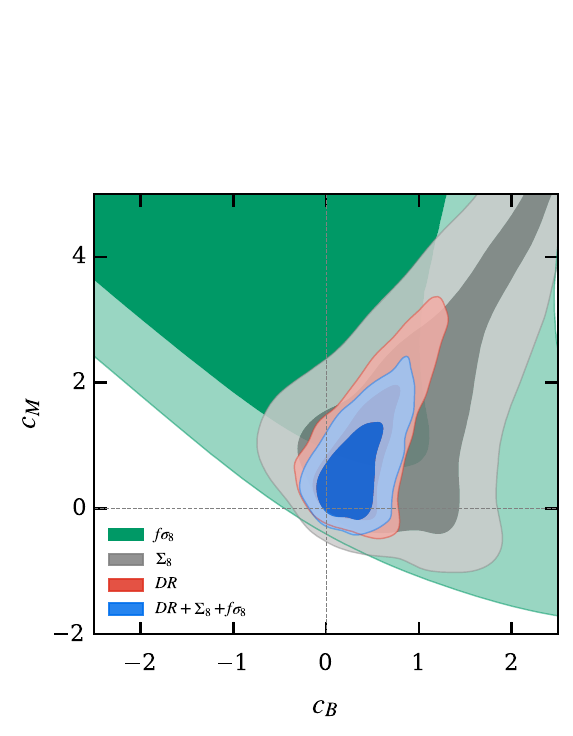}
\caption{Two-dimensional posterior contours for the three MG models considered in this work. From left to right, the panels show the results for the $(\mu_0,\eta_0)$, $(\mu_0,\Sigma_0)$, and $(c_{\rm M},c_{\rm B})$ parameterizations. The coloured contours compare the baseline joint constraints with those obtained from each probe individually, illustrating the synergy between $\mathit{DR}$ and other growth probes in reducing parameter degeneracies. In the $(\mu_0,\Sigma_0)$ case, the $f\sigma_8$ constraint is nearly orthogonal to the degeneracy direction preferred by $\mathit{DR}$ and $\Sigma_8$, and thus efficiently helps to break the $\mu_0$--$\Sigma_0$ degeneracy. In the EFT free-parameter case, single-probe constraints are generally weak, especially for $f\sigma_8$, whereas the joint analysis yields substantially tighter contours. We also note that $\mathit{DR}$ alone often provides stronger constraints than $\Sigma_8$ or $f\sigma_8$, consistent with its larger parameter sensitivities (Fig.~\ref{fig:sensitivity}). Contours correspond to the $68\%$ and $95\%$ credible regions.}
\label{fig:contour}
\end{figure*}
We use $\mathit{DR}$ measurements, $\Sigma_8$ constraints from CMB lensing tomography, and $f\sigma_8$ constraints from DESI full-shape and peculiar-velocity analyses. On the modeling side, we consider the $(\mu_0,\eta_0)$ and $(\mu_0,\Sigma_0)$ phenomenological parameterizations and the EFT $\alpha$-basis with parameters $(c_{\rm M},c_{\rm B})$.

We constrain the MG parameters using a Markov chain Monte Carlo (MCMC) approach and adopt a Gaussian likelihood,
\begin{equation}
\chi^{2}=\sum_s\sum_i
\frac{\left(D_i^s-T_i^s\right)^2}{(\sigma_i^s)^2},
\label{eq:chi2_def}
\end{equation}
where $s$ labels the datasets ($\mathit{DR}$, $\Sigma_8$, and $f\sigma_8$), and $i$ labels the individual redshift bins within each dataset. Here, $D_i^s$, $T_i^s$, and $\sigma_i^s$ denote the measured value, theoretical prediction, and reported $1\sigma$ uncertainty, respectively.

We ignore the cross-covariance between the different observational probes because the three observables are dominated by different sources of uncertainty: $\mathit{DR}$ by the low signal-to-noise of the ISW measurement, $\Sigma_8$ by CMB lensing reconstruction, and $f\sigma_8$ by RSD and peculiar velocity measurements. Where possible, we also select measurements with limited overlap in tracer samples and redshift coverage. Nevertheless, correlations between neighboring tomographic bins may remain. Since a complete covariance matrix for the compressed data vector is not available in a directly usable form, we adopt a diagonal covariance matrix and regard this as a limitation of the present analysis.

In addition, the compressed $\Sigma_8$ and $f\sigma_8$ measurements may retain some model dependence and omit correlations between MG and nuisance parameters. A full treatment of these effects would require reanalysis of the original likelihoods, which is beyond the scope of this work and is left for future study. To reduce sensitivity to this potential model dependence, we conservatively inflate the reported $\Sigma_8$ and $f\sigma_8$ uncertainties by $50\%$. For reference, we also present the results obtained without this error inflation in Appendix \ref{app:contours}.

We employ the MCMC sampler from the publicly available package \textsc{Cobaya} \cite{2021JCAP...05..057T} to perform likelihood sampling. Chain convergence is monitored with the generalized Gelman--Rubin statistic \cite{1992StaSc...7..457G,2013PhRvD..87j3529L}, and we require $R-1<0.01$ as our convergence criterion. To lessen sensitivity to initialization, we discard the first $30\%$ of each chain as burn-in. As indicated by Eq.~\ref{eq:DR_muSigma}, aside from the MG parameters, the $\mathit{DR}$ depends only on $\Omega_{\rm m}$. The value of $\Omega_{\rm m}$ is already tightly constrained by a range of expansion-history measurements, including CMB, BAO, and SN~Ia. We therefore adopt a Gaussian prior on $\Omega_{\rm m}$ centred on the \textit{Planck} 2018 best-fit value\cite{2020A&A...641A...1P}. Accordingly, in our parameter inference we vary only the MG parameters together with $\sigma_8(0)$, while fixing the remaining cosmological parameters. Table~\ref{tab:priors} summarizes the priors adopted for these parameters.

\begin{table}
    \centering
    \begin{tabular}{c|c}
    \hline
      parameter &prior\\
      \hline
      $\sigma_8(0)$ & $\mathcal{U}[0,2]$\\
      $\Omega_{m,0}$ & $\mathcal{N}(0.3153,0.0073^2)$\\
      $\mu_0$ & $\mathcal{U}[-10,10]$\\
      $\Sigma_0$ & $\mathcal{U}[-10,10]$\\
      $\eta_0$ & $\mathcal{U}[-10,10]$\\
      $c_{\rm M}$ & $\mathcal{U}[-10,10]$\\
      $c_{\rm B}$ & $\mathcal{U}[-10,10]$\\
    \hline
    \end{tabular}
    \caption{Parameters and priors used in the analysis. The prior on $\Omega_{m,0}$ is adopted from \cite{2020A&A...641A...1P}.}
    \label{tab:priors}
\end{table}

\subsection{Constraints on phenomenological parameterizations}\label{sec:res_phen}

We first present constraints on our baseline (DR+$\Sigma_8$+$f\sigma_8$) phenomenological parameterizations.
Marginalizing over $\sigma_8(0)$, we obtain
\begin{equation}
\mu_0 = 0.06^{+0.17}_{-0.23},\qquad
\eta_0 = -0.03^{+0.36}_{-0.46},
\end{equation}
for the $(\mu_0,\eta_0)$ framework, and
\begin{equation}
\mu_0 = 0.09 \pm 0.35,\qquad
\Sigma_0 = 0.01 \pm 0.06,
\end{equation}
for the $(\mu_0,\Sigma_0)$ framework (quoted at $68\%$ confidence). Both parameterizations are consistent with the GR expectation ($\mu_0=\eta_0=\Sigma_0=0$). The corresponding posterior contours are shown in Figure~\ref{fig:contour}, and the numerical results are summarized in Table~\ref{tab:results_summary}. 

Figure~\ref{fig:contour} compares the baseline joint constraints with those obtained from each probe individually. 
Overall, the combined data vector yields tighter posteriors than any single observable, providing a direct visual demonstration of the synergy between $\mathit{DR}$ and the other growth probes. 
Figure~\ref{fig:sensitivity} shows that these probes exhibit different sensitivity amplitudes and sign patterns across the MG parameters. Consequently, they are sensitive to different aspects of the parameter space, so combining $\mathit{DR}$, $\Sigma_8$, and $f\sigma_8$ reduces degeneracies and improves the overall precision.
This complementarity is particularly clear in the $(\mu_0,\Sigma_0)$ plane. The $f\sigma_8$-only constraint follows a degeneracy direction that is nearly orthogonal to that preferred by $\mathit{DR}$ and $\Sigma_8$, so adding $f\sigma_8$ efficiently sharpens the joint constraint by anchoring the $\mu_0$ direction.

To further quantify this complementarity, we compare the three two-probe combinations ($\mathit{DR}$+$\Sigma_8$, $\mathit{DR}$+$f\sigma_8$, and $\Sigma_8+f\sigma_8$; see Table~\ref{tab:results_summary}, with the corresponding contours shown in Appendix~\ref{app:contours}).

In the $(\mu_0,\Sigma_0)$ framework, the $\mathit{DR}$+$\Sigma_8$ combination yields very weak constraints,
\begin{equation}
\mu_0 = 3.0 \pm 2.7,\qquad
\Sigma_0 = -0.23^{+0.15}_{-0.29},
\end{equation}
reflecting the fact that both $\mathit{DR}$ and $\Sigma_8$ respond only weakly to $\mu_0$ (Figure~\ref{fig:sensitivity}), which leaves a pronounced $\mu_0$--$\Sigma_0$ degeneracy. This is also evident in Figure~\ref{fig:contour}, where the $\mathit{DR}$-only and $\Sigma_8$-only contours follow nearly the same degeneracy direction.

Including $f\sigma_8$ changes the picture substantially. Because $f\sigma_8$ provides direct growth rate information that anchors $\mu_0$ in our framework, the two-probe combinations that include $f\sigma_8$ become much more informative:
\begin{equation}
(\mathit{DR}+f\sigma_8):\ \ \mu_0=-0.04\pm0.57,\ \ \Sigma_0=0.030\pm0.087,
\end{equation}
\begin{equation}
(\Sigma_8+f\sigma_8):\ \ \mu_0=0.1\pm0.44,\ \ \Sigma_0=0.01^{+0.12}_{-0.15}.
\end{equation}
In particular, once $\mu_0$ is better localized by $f\sigma_8$, DR provides complementary sensitivity that helps constrain $\Sigma_0$ through potential evolution information. The baseline combination then yields the most informative overall posterior by combining the DR-driven potential evolution information with complementary growth constraints.
For the $(\mu_0,\eta_0)$ framework, the contrast among two-probe combinations is less extreme. The DR+$\Sigma_8$ constraints remain relatively informative,
\begin{equation}
(\mathit{DR}+\Sigma_8):\ \ \mu_0 = 0.06^{+0.20}_{-0.28},\ \ \eta_0 = 0.02^{+0.43}_{-0.62},
\end{equation}
consistent with $\mathit{DR}$ being relatively sensitive to both $\mu_0$ and $\eta_0$ in this parameterization (Figure~\ref{fig:sensitivity}). The other two-probe combinations yield broadly consistent results,
\begin{equation}
(\mathit{DR}+f\sigma_8):\ \ \mu_0 = 0.02^{+0.20}_{-0.29},\ \ \eta_0 = 0.10^{+0.48}_{-0.65},
\end{equation}
\begin{equation}
(\Sigma_8+f\sigma_8):\ \ \mu_0 = 0.07^{+0.22}_{-0.39},\ \ \eta_0 = -0.02^{+0.48}_{-0.63},
\end{equation}
while the baseline data vector still provides the most constraining and stable posteriors overall. Taken together, these comparisons show that $\mathit{DR}$, $\Sigma_8$, and $f\sigma_8$ provide complementary growth information, and their combination yields the strongest phenomenological constraints by reducing parameter degeneracies.

\subsection{Constraints on EFT parameterizations}\label{sec:res_eft}

We next present constraints in the EFT of DE $\alpha$-basis. For the free $(c_{\rm M},c_{\rm B})$ case, we obtain
\begin{equation}
c_{\rm M}=0.64^{+0.32}_{-0.72},\qquad
c_{\rm B}=0.31^{+0.19}_{-0.29},
\end{equation}
with the GR limit $(c_{\rm M},c_{\rm B})=(0,0)$ remaining consistent at the $95\%$ confidence level (see Fig.~\ref{fig:contour}). Although the GR point lies outside the $68\%$ contour in the free EFT case, this does not conflict with the phenomenological constraints. As shown by Eqs.~\eqref{eq:mu_alpha_qsa} and \eqref{eq:sigma_alpha_qsa}, $\mu$ and $\Sigma$ are nonlinear, redshift-dependent functions of $(c_{\rm M},c_{\rm B})$, so the EFT model is not a simple reparameterization of $(\mu_0,\Sigma_0)$. The marginalized credible regions in the two parameter spaces therefore need not map onto one another directly. Since the GR point remains within the $95\%$ contour, we find no statistically significant evidence for a departure from GR.

This conclusion is qualitatively consistent with the recent multiprobe EFTofDE analysis of \cite{2026PhRvD.113f3555L}, who combined  the BOSS full-shape galaxy power spectrum and bispectrum with  angular galaxy-clustering and CMB cross-correlation measurements,  together with CMB, BAO, and supernova observations. Their constraints  on $(c_{\rm B},c_{\rm M})$ were likewise found to be compatible with GR at approximately the $2\sigma$ level.

For a more direct comparison, \cite{2025JCAP...09..053I} combined DESI full-shape clustering and DESI BAO measurements with CMB data from \textit{Planck} PR4 and ACT DR6 and the DES-SN5YR supernova dataset. Compared with their results, our marginalized uncertainties are smaller by approximately a factor of $\sim2$ for $c_{\rm M}$ and $\sim1.5$ for $c_{\rm B}$. A plausible origin of this improvement is that large values of $\alpha_{\rm M}$ or $\alpha_{\rm B}$ can modify the late-time evolution of the gravitational potentials \cite{2017JCAP...08..019Z,2019PhRvD..99j3502N}, while $\mathit{DR}$ is explicitly constructed to trace their temporal evolution. Consequently, $\mathit{DR}$ provides a particularly relevant probe of late-time EFT departures from GR. We note that differences in the adopted data combinations and priors may also contribute to the quantitative comparison with the DESI analysis.

Figure~\ref{fig:contour} also compares the baseline joint constraints with those obtained from each probe individually. The single-probe constraints are generally weak, particularly for $f\sigma_8$, whereas the joint analysis yields substantially tighter contours. Importantly, even when used alone, $\mathit{DR}$ delivers significantly tighter constraints than $\Sigma_8$ or $f\sigma_8$, consistent with its larger sensitivities to $(c_{\rm M},c_{\rm B})$ in Fig.~\ref{fig:sensitivity}. This highlights that the dominant gain in the free EFT case comes from incorporating $\mathit{DR}$ as a direct probe, with the growth probes providing complementary information that further stabilizes and tightens the joint posterior.

This DR-driven synergy is also evident from the two-probe combinations (Table~\ref{tab:results_summary}; contours in Appendix~\ref{app:contours}). $\Sigma_8+f\sigma_8$ combination yields the weakest constraints,
\begin{equation}
(\Sigma_8+f\sigma_8):\ \ c_{\rm M}=1.66^{+0.51}_{-1.9},\qquad c_{\rm B}=0.84^{+0.49}_{-0.92},
\end{equation}
whereas including $\mathit{DR}$ leads to a marked improvement,
\begin{equation}
(\mathrm{DR}+\Sigma_8):\ \ c_{\rm M}=0.71^{+0.34}_{-0.84},\qquad c_{\rm B}=0.38^{+0.23}_{-0.31},
\end{equation}
\begin{equation}
(\mathrm{DR}+f\sigma_8):\ \ c_{\rm M}=0.78^{+0.38}_{-0.83},\qquad c_{\rm B}=0.36^{+0.23}_{-0.30}.
\end{equation}
The baseline combination yields the tightest overall constraints, consistent with $\mathit{DR}$ providing the key potential-evolution information and the growth probes adding complementary constraints that help reduce residual degeneracies.


For the no-braiding subclass ($\alpha_{\rm B}=0$), the stability requirements enforce $\alpha_{\rm M}>0$, and we obtain an upper limit $c_{\rm M}<0.479$.
For the no-slip case ($\alpha_{\rm B}=-2\alpha_{\rm M}$), we find $c_{\rm M}=-0.07\pm0.15$.
In both subclasses, the resulting precision is comparable to that achieved by DESI \cite{2025JCAP...09..053I}. This is expected, since in these cases the $c_{\rm M}$--$c_{\rm B}$ degeneracy is effectively reduced by imposing the parameter relations a priori; consequently, the additional constraining power enabled by $\mathit{DR}$ in the free $(c_{\rm M},c_{\rm B})$ analysis is diminished, and no substantial improvement in precision is obtained.

\begin{table*}[t]
\centering
\caption{Summary of constraints on phenomenological MG parameterizations and EFT $\alpha$-basis models. Uncertainties correspond to $68\%$ credible intervals. The shorthand ``$\Omega_{\rm DE}$'' denotes the time dependence $P_{\rm MG}(a)=1+P_{{\rm MG},0}\,\Omega_{\rm DE}(a)/\Omega_{\rm DE}(0)$.}
\label{tab:results_summary}
\setlength{\tabcolsep}{6.5pt}
\renewcommand{\arraystretch}{1.15}
\begin{tabular}{lllc c}
\toprule
Model & Time dep. & Data combo & Parameter 1 & Parameter 2 \\
\midrule
\multicolumn{5}{l}{\textbf{Phenomenological parameterizations (baseline data vector: DR+$\Sigma_8$+$f\sigma_8$)}}\\
$(\mu_0,\Sigma_0)$ & $\Omega_{\rm DE}$ & Baseline & $\mu_0=0.09\pm0.35$          & $\Sigma_0=0.01\pm0.06$ \\
$(\mu_0,\eta_0)$   & $\Omega_{\rm DE}$ & Baseline & $\mu_0=0.06^{+0.17}_{-0.23}$ & $\eta_0=-0.03^{+0.36}_{-0.46}$ \\
\midrule
\multicolumn{5}{l}{\textbf{EFT of DE in the $\alpha$-basis}}\\
$(c_{\rm M},c_{\rm B})$ & $\Omega_{\rm DE}$ & Baseline & $c_{\rm M}=0.64^{+0.32}_{-0.72}$ & $c_{\rm B}=0.31^{+0.19}_{-0.29}$ \\
($\alpha_{\rm B}=0$) & $\Omega_{\rm DE}$ & Baseline & $c_{\rm M}<0.479$ & $c_{\rm B}=0$ \\
($\alpha_{\rm B}=-2\alpha_{\rm M}$) & $\Omega_{\rm DE}$ & Baseline & $c_{\rm M}=-0.07\pm0.15$ & $c_{\rm B}=-2c_{\rm M}$ \\
\midrule
\multicolumn{5}{l}{\textbf{Restricted data combinations (for comparison)}}\\
$(\mu_0,\Sigma_0)$ & $\Omega_{\rm DE}$ & DR+$\Sigma_8$ & $\mu_0=3.0\pm2.7$            & $\Sigma_0=-0.23^{+0.15}_{-0.29}$ \\
$(\mu_0,\Sigma_0)$ & $\Omega_{\rm DE}$ & DR+$f\sigma_8$ & $\mu_0=-0.04\pm0.57$         & $\Sigma_0=0.030\pm0.087$ \\
$(\mu_0,\Sigma_0)$ & $\Omega_{\rm DE}$ & $\Sigma_8$+$f\sigma_8$ & $\mu_0=0.1\pm0.44$         & $\Sigma_0=0.01^{+0.12}_{-0.15}$ \\
\addlinespace
$(\mu_0,\eta_0)$   & $\Omega_{\rm DE}$ & DR+$\Sigma_8$ & $\mu_0=0.06^{+0.20}_{-0.28}$ & $\eta_0=0.02^{+0.43}_{-0.62}$ \\
$(\mu_0,\eta_0)$   & $\Omega_{\rm DE}$ & DR+$f\sigma_8$ & $\mu_0=0.02^{+0.20}_{-0.29}$ & $\eta_0=0.10^{+0.48}_{-0.65}$ \\
$(\mu_0,\eta_0)$   & $\Omega_{\rm DE}$ & $\Sigma_8$+$f\sigma_8$ & $\mu_0=0.07^{+0.22}_{-0.39}$ & $\eta_0=-0.02^{+0.48}_{-0.63}$ \\
\addlinespace
$(c_{\rm M},c_{\rm B})$ & $\Omega_{\rm DE}$ & DR+$\Sigma_8$ & $c_{\rm M}=0.71^{+0.34}_{-0.84}$ & $c_{\rm B}=0.38^{+0.23}_{-0.31}$ \\
$(c_{\rm M},c_{\rm B})$ & $\Omega_{\rm DE}$ & DR+$f\sigma_8$ & $c_{\rm M}=0.78^{+0.38}_{-0.83}$ & $c_{\rm B}=0.36^{+0.23}_{-0.30}$ \\
$(c_{\rm M},c_{\rm B})$ & $\Omega_{\rm DE}$ & $\Sigma_8$+$f\sigma_8$ & $c_{\rm M}=1.66^{+0.51}_{-1.9}$ & $c_{\rm B}=0.84^{+0.49}_{-0.92}$ \\
\bottomrule
\end{tabular}
\end{table*}

\section{Summary and Outlook}\label{sec:summary}

We have tested gravity by exploiting the synergy among three growth-related observables: the gravitational potential decay rate ($\mathit{DR}$), the lensing-inferred growth amplitude $\Sigma_8$, and the growth rate measurement $f\sigma_8$. These probes respond to MG parameters with different sensitivity amplitudes and sign patterns, and their combination therefore reduces parameter degeneracies relative to any single probe or two-probe subset.

Within phenomenological parameterizations, our baseline constraints are consistent with GR. Relative to the  combination $\Sigma_8+f\sigma_8$, including $\mathit{DR}$ tightens the $(\mu_0,\Sigma_0)$ constraints mainly through an improvement of about a factor of $\sim2$ on $\Sigma_0$, and improves the $(\mu_0,\eta_0)$ constraints on both parameters by roughly a factor of $\sim1.5$ (see Table~\ref{tab:results_summary}). In the EFT $\alpha$-basis, we obtain constraints consistent with GR and find that the free $(c_{\rm M},c_{\rm B})$ uncertainties are about a factor of $\sim2$ smaller than those reported in \citep{2025JCAP...09..053I}, which combined DESI full-shape and BAO measurements with DES-SN5YR and CMB data. This improvement is consistent with $\mathit{DR}$ providing direct sensitivity to late-time potential evolution, which is particularly informative for EFT-based tests.

Looking ahead, more precise tomographic $\mathit{DR}$ measurements with higher signal-to-noise and extended redshift coverage will further enhance this complementarity-based approach, especially for multi-parameter EFT tests. Improved public covariance information for multi-probe combinations will also enable more rigorous joint analyses and sharpen gravity tests in the late Universe.

\appendix
\renewcommand{\thefigure}{A\arabic{figure}}
\setcounter{figure}{0}
\section{Additional results and posterior contours}\label{app:contours}

\begin{figure*}[t]
\centering
\includegraphics[width=0.3\textwidth]{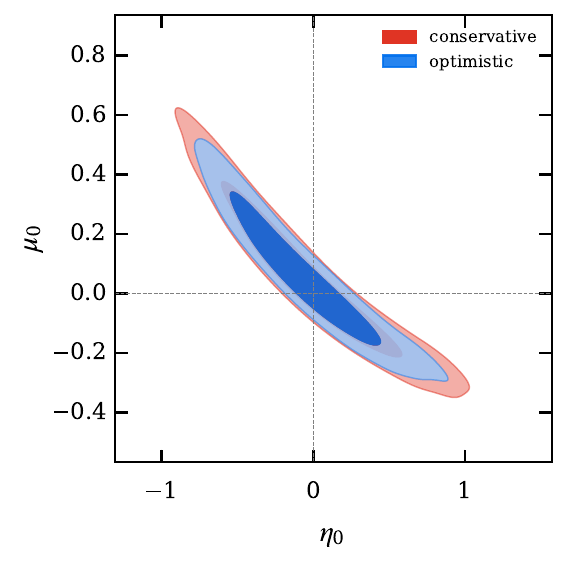}\hfill
\includegraphics[width=0.3\textwidth]{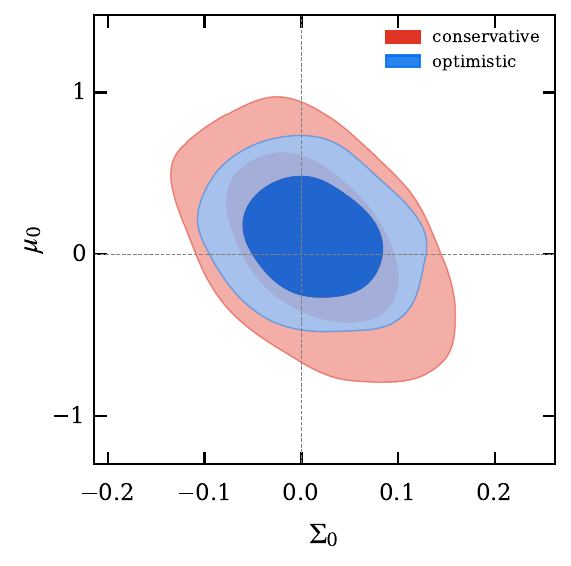}\hfill
\includegraphics[width=0.3\textwidth]{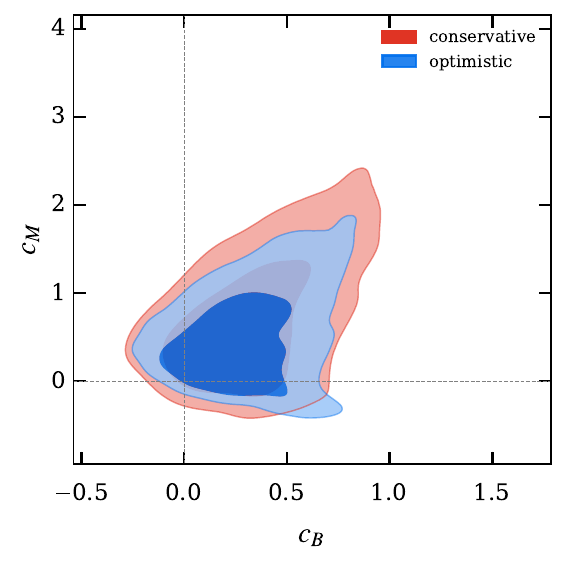}
\caption{Constraints obtained with (conservative; red) and without (optimistic; blue) the $50\%$ error inflation applied to the $\Sigma_8$ and $f\sigma_8$ measurements. The optimistic case yields tighter contours, while the conservative choice provides a more cautious assessment; the qualitative conclusions, including consistency with GR, remain unchanged. All calculations are performed for the baseline setup. From left to right, the panels show the $(\mu_0,\eta_0)$, $(\mu_0,\Sigma_0)$, and $(c_{\rm M},c_{\rm B})$ parameterizations. Contours indicate the $68\%$ and $95\%$ credible regions.}
\label{fig:contour_two_compare_baseline}
\end{figure*}

In this appendix we present supplementary results and contour comparisons. We first repeat the baseline analysis without inflating the quoted uncertainties of $\Sigma_8$ and $f\sigma_8$ (``optimistic'' case). The resulting constraints are
\begin{equation}
\mu_0 = 0.10\pm0.25,\qquad \Sigma_0=0.012\pm0.048,
\end{equation}
\begin{equation}
\mu_0 = 0.06^{+0.15}_{-0.20},\qquad \eta_0=-0.06^{+0.30}_{-0.38},
\end{equation}
and
\begin{equation}
c_{\rm M}=0.44^{+0.28}_{-0.55},\qquad c_{\rm B}=0.28^{+0.19}_{-0.25},
\end{equation}
for the $(\mu_0,\Sigma_0)$, $(\mu_0,\eta_0)$, and $(c_{\rm M},c_{\rm B})$ parameterizations, respectively. The corresponding posterior contours, together with those from our conservative baseline choice (with $50\%$ error inflation), are shown in Fig.~\ref{fig:contour_two_compare_baseline}. As expected, removing the error inflation yields tighter constraints, while the qualitative conclusions, including consistency with GR and the overall degeneracy directions, remain unchanged.

\begin{figure*}[t]
\centering
\includegraphics[width=0.3\textwidth]{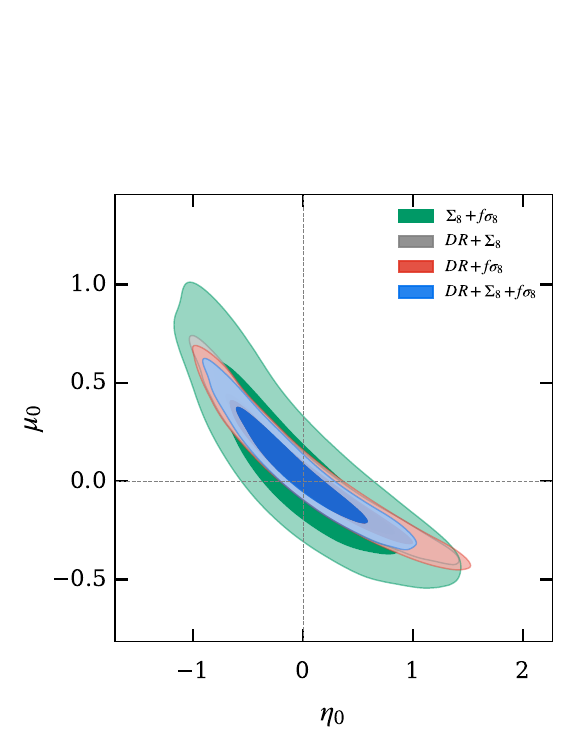}\hfill
\includegraphics[width=0.3\textwidth]{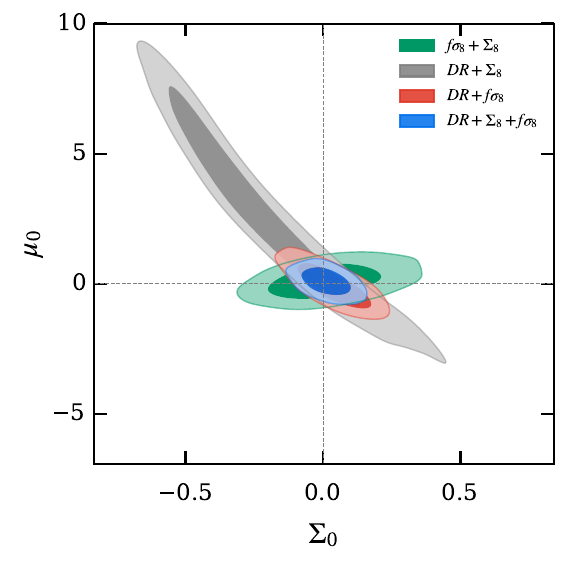}\hfill
\includegraphics[width=0.3\textwidth]{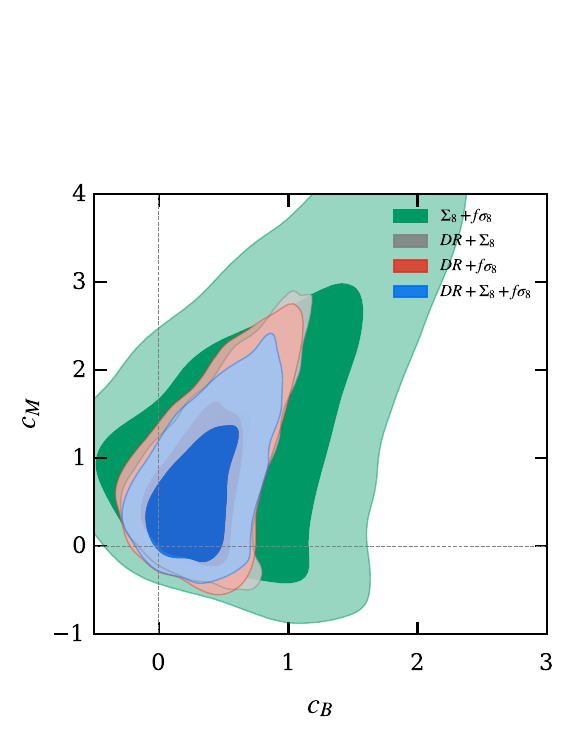}
\caption{Two-dimensional posterior contours for the three MG models considered in this work, comparing different data combinations. From left to right, the panels show the $(\mu_0,\eta_0)$, $(\mu_0,\Sigma_0)$, and $(c_{\rm M},c_{\rm B})$ parameterizations. In the $(\mu_0,\Sigma_0)$ case, the DR+$\Sigma_8$ combination yields relatively weak constraints because both probes have limited sensitivity to $\mu_0$, whereas adding $f\sigma_8$ substantially improves the constraints by providing complementary information on $\mu_0$. Contours indicate the $68\%$ and $95\%$ credible regions; dashed lines mark the GR values.}
\label{fig:contour_two_compare}
\end{figure*}

\begin{figure*}[t]
\centering
\includegraphics[width=0.45\textwidth]{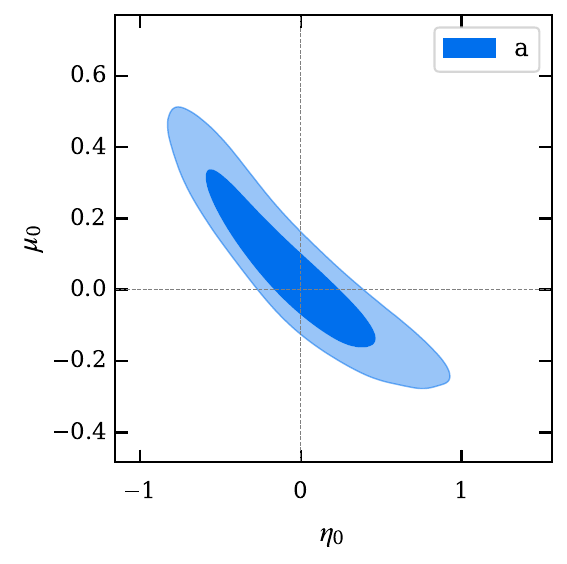}\hfill
\includegraphics[width=0.45\textwidth]{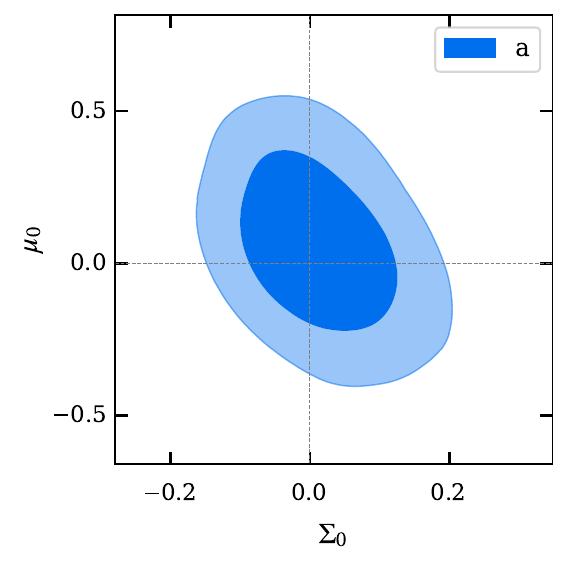}\hfill
\includegraphics[width=0.45\textwidth]{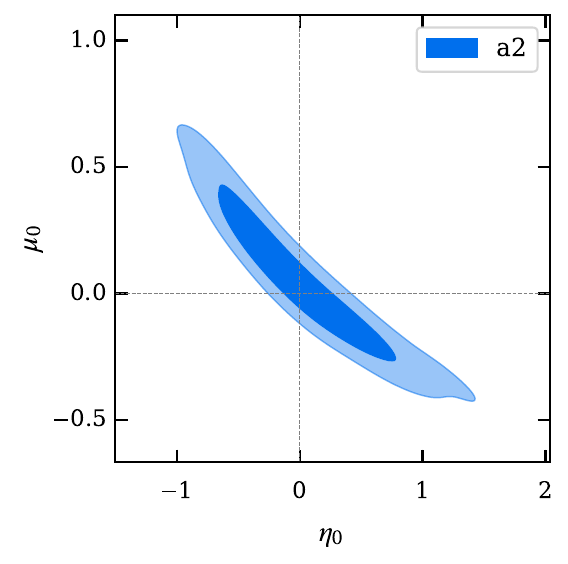}\hfill
\includegraphics[width=0.45\textwidth]{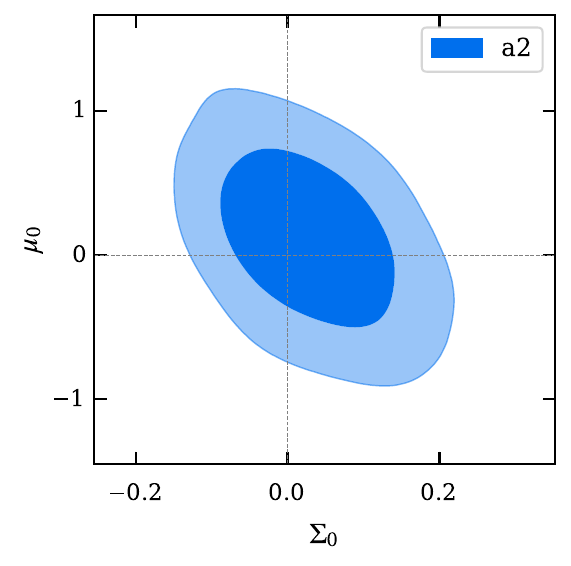}
\caption{Posterior contours for the alternative phenomenological time parameterizations. The top row shows the results for $P_{\rm MG}(a)=1+P_{{\rm MG},0}a$, while the bottom row shows the results for $P_{\rm MG}(a)=1+P_{{\rm MG},0}a^{2}$.}
\label{fig:contour_a_a2}
\end{figure*}

We then provide additional contour comparisons between the baseline data vector and three restricted combinations,$\ \Sigma_8$+$f\sigma_8$, DR+$\Sigma_8$ and DR+$f\sigma_8$, shown in Fig.~\ref{fig:contour_two_compare}. 
Finally, we test two alternative time-dependence prescriptions for the phenomenological parameterization, namely $P_{\rm MG}(a)=1+P_{{\rm MG},0}\,a$ and $P_{\rm MG}(a)=1+P_{{\rm MG},0}\,a^{2}$. We obtain
\begin{equation}
\mu_0 = 0.06^{+0.14}_{-0.19},\qquad
\eta_0 = -0.06^{+0.31}_{-0.40}\,,
\end{equation}
and
\begin{equation}
\mu_0 = 0.06^{+0.21}_{-0.26},\qquad 
\eta_0 = 0.01^{+0.41}_{-0.57}\,,
\end{equation}
for the $(\mu_0,\eta_0)$ framework and
\begin{equation}
\mu_0 = 0.06\pm 0.19,\qquad
\Sigma_0 = 0.01^{+0.07}_{-0.08}\,,
\end{equation}
and
\begin{equation}
\mu_0 = 0.11\pm 0.41,\qquad
\Sigma_0 = 0.03\pm 0.08\,,
\end{equation}
for the $(\mu_0,\Sigma_0)$ framework, respectively. The corresponding posterior contours are shown in Fig.~\ref{fig:contour_a_a2} for comparison with the baseline constraints in the main text. We find that these constraints remain consistent with GR even when adopting different parameterizations.

\acknowledgments
Pengjie Zhang and Shang Li are supported by  the National Key R\&D Program of China (2023YFA1607800, 2023YFA1607801).
Fuyu Dong is supported by  the National Natural Science Foundation of China (grant No.12303003).

\bibliographystyle{JHEP}
\bibliography{biblio.bib}

@ARTICLE{2019ApJS..242....8Z,
       author = {{Zou}, Hu and {Gao}, Jinghua and {Zhou}, Xu and {Kong}, Xu},
        title = "{Photometric Redshifts and Stellar Masses for Galaxies from the DESI Legacy Imaging Surveys}",
      journal = {\apjs},
         year = 2019,
        month = may,
       volume = {242},
       number = {1},
          eid = {8},
        pages = {8},
          doi = {10.3847/1538-4365/ab1847},
       adsurl = {https://ui.adsabs.harvard.edu/abs/2019ApJS..242....8Z}
}

@ARTICLE{2022ApJ...938...72D,
       author = {{Dong}, Fuyu and {Zhang}, Pengjie and {Sun}, Zeyang and {Park}, Changbom},
        title = "{The First Direct Measurement of Gravitational Potential Decay Rate at Cosmological Scales and Improved Dark Energy Constraint}",
      journal = {\apj},
         year = 2022,
        month = oct,
       volume = {938},
       number = {1},
          eid = {72},
        pages = {72},
          doi = {10.3847/1538-4357/ac905b},
archivePrefix = {arXiv},
       eprint = {2206.04917},
 primaryClass = {astro-ph.CO},
       adsurl = {https://ui.adsabs.harvard.edu/abs/2022ApJ...938...72D}
}

@ARTICLE{2025ApJ...982...99D,
       author = {{Dong}, Fuyu and {Zhang}, Pengjie and {Xu}, Haojie and {Qin}, Jian},
        title = "{The Direct Measurement of Gravitational Potential Decay Rate at Cosmological Scales. II. Improved Dark Energy Constraint from z {\ensuremath{\leq}} 1.4}",
      journal = {\apj},
         year = 2025,
        month = apr,
       volume = {982},
       number = {2},
          eid = {99},
        pages = {99},
          doi = {10.3847/1538-4357/adb585},
archivePrefix = {arXiv},
       eprint = {2411.12594},
 primaryClass = {astro-ph.CO},
       adsurl = {https://ui.adsabs.harvard.edu/abs/2025ApJ...982...99D}
}

@ARTICLE{2006ApJ...647...55Z,
       author = {{Zhang}, Pengjie},
        title = "{Isolating the Decay Rate of Cosmological Gravitational Potential}",
      journal = {\apj},
         year = 2006,
        month = aug,
       volume = {647},
       number = {1},
        pages = {55-59},
          doi = {10.1086/505297},
archivePrefix = {arXiv},
       eprint = {astro-ph/0512422},
 primaryClass = {astro-ph},
       adsurl = {https://ui.adsabs.harvard.edu/abs/2006ApJ...647...55Z}
}

@ARTICLE{2025JCAP...10..077D,
       author = {{de Belsunce}, R. and {Krolewski}, A. and {Chaussidon}, E. and {Ferraro}, S. and {Farren}, G. and {Hadzhiyska}, B. and {Tamone}, A. and {Chiarenza}, S. and {Sailer}, N. and {Ravoux}, C. and et al.},
        title = "{Cosmology from Planck CMB lensing and DESI DR1 quasar tomography}",
      journal = {\jcap},
         year = 2025,
        month = oct,
       volume = {2025},
       number = {10},
          eid = {077},
        pages = {077},
          doi = {10.1088/1475-7516/2025/10/077},
archivePrefix = {arXiv},
       eprint = {2506.22416},
 primaryClass = {astro-ph.CO},
       adsurl = {https://ui.adsabs.harvard.edu/abs/2025JCAP...10..077D}
}

@ARTICLE{2025JCAP...06..008S,
       author = {{Sailer}, Noah and {Kim}, Joshua and {Ferraro}, Simone and {Madhavacheril}, Mathew S. and {White}, Martin and {Abril-Cabezas}, Irene and {Aguilar}, Jessica Nicole and {Ahlen}, Steven and {Richard Bond}, J. and {Brooks}, David and et al.},
        title = "{Cosmological constraints from the cross-correlation of DESI Luminous Red Galaxies with CMB lensing from Planck PR4 and ACT DR6}",
      journal = {\jcap},
         year = 2025,
        month = jun,
       volume = {2025},
       number = {6},
          eid = {008},
        pages = {008},
          doi = {10.1088/1475-7516/2025/06/008},
archivePrefix = {arXiv},
       eprint = {2407.04607},
 primaryClass = {astro-ph.CO},
       adsurl = {https://ui.adsabs.harvard.edu/abs/2025JCAP...06..008S}
}

@ARTICLE{2025A&A...696A...5A,
       author = {{Artis}, E. and {Bulbul}, E. and {Grandis}, S. and {Ghirardini}, V. and {Clerc}, N. and {Seppi}, R. and {Comparat}, J. and {Cataneo}, M. and {von der Linden}, A. and {Bahar}, Y.~E. and et al.},
        title = "{The SRG/eROSITA All-Sky Survey: Constraints on the structure growth from cluster number counts}",
      journal = {\aap},
         year = 2025,
        month = apr,
       volume = {696},
          eid = {A5},
        pages = {A5},
          doi = {10.1051/0004-6361/202452584},
archivePrefix = {arXiv},
       eprint = {2410.09499},
 primaryClass = {astro-ph.CO},
       adsurl = {https://ui.adsabs.harvard.edu/abs/2025A&A...696A...5A}
}

@ARTICLE{2023PhRvD.107h3504A,
       author = {{Abbott}, T.~M.~C. and {Aguena}, M. and {Alarcon}, A. and {Alves}, O. and {Amon}, A. and {Andrade-Oliveira}, F. and {Annis}, J. and {Avila}, S. and {Bacon}, D. and {Baxter}, E. and et al.},
        title = "{Dark Energy Survey Year 3 results: Constraints on extensions to {\ensuremath{\Lambda}} CDM with weak lensing and galaxy clustering}",
      journal = {\prd},
         year = 2023,
        month = apr,
       volume = {107},
       number = {8},
          eid = {083504},
        pages = {083504},
          doi = {10.1103/PhysRevD.107.083504},
archivePrefix = {arXiv},
       eprint = {2207.05766},
 primaryClass = {astro-ph.CO},
       adsurl = {https://ui.adsabs.harvard.edu/abs/2023PhRvD.107h3504A}
}

@ARTICLE{2025arXiv251009563R,
       author = {{Rubiola}, Andrea and {Zennaro}, Matteo and {Garc{\'\i}a-Garc{\'\i}a}, Carlos and {Alonso}, David},
        title = "{Low-redshift constraints on structure growth from CMB lensing tomography}",
      journal = {arXiv e-prints},
         year = 2025,
        month = oct,
          eid = {arXiv:2510.09563},
        pages = {arXiv:2510.09563},
          doi = {10.48550/arXiv.2510.09563},
archivePrefix = {arXiv},
       eprint = {2510.09563},
 primaryClass = {astro-ph.CO},
       adsurl = {https://ui.adsabs.harvard.edu/abs/2025arXiv251009563R}
}

@ARTICLE{2024JCAP...06..012P,
       author = {{Piccirilli}, Giulia and {Fabbian}, Giulio and {Alonso}, David and {Storey-Fisher}, Kate and {Carron}, Julien and {Lewis}, Antony and {Garc{\'\i}a-Garc{\'\i}a}, Carlos},
        title = "{Growth history and quasar bias evolution at z < 3 from Quaia}",
      journal = {\jcap},
         year = 2024,
        month = jun,
       volume = {2024},
       number = {6},
          eid = {012},
        pages = {012},
          doi = {10.1088/1475-7516/2024/06/012},
archivePrefix = {arXiv},
       eprint = {2402.05761},
 primaryClass = {astro-ph.CO},
       adsurl = {https://ui.adsabs.harvard.edu/abs/2024JCAP...06..012P}
}

@ARTICLE{2025JCAP...07..028A,
       author = {{Adame}, A.~G. and {Aguilar}, J. and {Ahlen}, S. and {Alam}, S. and {Alexander}, D.~M. and {Allende Prieto}, C. and {Alvarez}, M. and {Alves}, O. and {Anand}, A. and {Andrade}, U. and et al.},
        title = "{DESI 2024 VII: cosmological constraints from the full-shape modeling of clustering measurements}",
      journal = {\jcap},
         year = 2025,
        month = jul,
       volume = {2025},
       number = {7},
          eid = {028},
        pages = {028},
          doi = {10.1088/1475-7516/2025/07/028},
archivePrefix = {arXiv},
       eprint = {2411.12022},
 primaryClass = {astro-ph.CO},
       adsurl = {https://ui.adsabs.harvard.edu/abs/2025JCAP...07..028A}
}

@ARTICLE{2025JCAP...09..008A,
       author = {{Adame}, A.~G. and {Aguilar}, J. and {Ahlen}, S. and {Alam}, S. and {Alexander}, D.~M. and {Alvarez}, M. and {Alves}, O. and {Anand}, A. and {Andrade}, U. and {Armengaud}, E. and et al.},
        title = "{DESI 2024 V: Full-Shape galaxy clustering from galaxies and quasars}",
      journal = {\jcap},
         year = 2025,
        month = sep,
       volume = {2025},
       number = {9},
          eid = {008},
        pages = {008},
          doi = {10.1088/1475-7516/2025/09/008},
archivePrefix = {arXiv},
       eprint = {2411.12021},
 primaryClass = {astro-ph.CO},
       adsurl = {https://ui.adsabs.harvard.edu/abs/2025JCAP...09..008A}
}

@ARTICLE{2025arXiv251203231Q,
       author = {{Qin}, F. and {Blake}, C. and {Howlett}, C. and {Turner}, R.~J. and {Lodha}, K. and {Bautista}, J. and {Lai}, Y. and {Amsellem}, A.~J. and {Aguilar}, J. and {Ahlen}, S. and et al.},
        title = "{The DESI DR1 Peculiar Velocity Survey: Growth Rate Measurements from the Galaxy Power Spectrum}",
      journal = {arXiv e-prints},
         year = 2025,
        month = dec,
          eid = {arXiv:2512.03231},
        pages = {arXiv:2512.03231},
          doi = {10.48550/arXiv.2512.03231},
archivePrefix = {arXiv},
       eprint = {2512.03231},
 primaryClass = {astro-ph.CO},
       adsurl = {https://ui.adsabs.harvard.edu/abs/2025arXiv251203231Q}
}

@ARTICLE{2025arXiv251203229L,
       author = {{Lai}, Y. and {Howlett}, C. and {Aguilar}, J. and {Ahlen}, S. and {Amsellem}, A.~J. and {Bautista}, J. and {BenZvi}, S. and {Bianchi}, D. and {Blake}, C. and {Brooks}, D. and et al.},
        title = "{The DESI DR1 Peculiar Velocity Survey: growth rate measurements from the maximum likelihood fields method}",
      journal = {arXiv e-prints},
         year = 2025,
        month = dec,
          eid = {arXiv:2512.03229},
        pages = {arXiv:2512.03229},
          doi = {10.48550/arXiv.2512.03229},
archivePrefix = {arXiv},
       eprint = {2512.03229},
 primaryClass = {astro-ph.CO},
       adsurl = {https://ui.adsabs.harvard.edu/abs/2025arXiv251203229L}
}

@ARTICLE{2025arXiv251203230T,
       author = {{Turner}, R.~J. and {Blake}, C. and {Qin}, F. and {Aguilar}, J. and {Ahlen}, S. and {Amsellem}, A.~J. and {Bautista}, J. and {BenZvi}, S. and {Bianchi}, D. and {Brooks}, D. and et al.},
        title = "{The DESI DR1 Peculiar Velocity Survey: growth rate measurements from galaxy and momentum correlation functions}",
      journal = {arXiv e-prints},
         year = 2025,
        month = dec,
          eid = {arXiv:2512.03230},
        pages = {arXiv:2512.03230},
          doi = {10.48550/arXiv.2512.03230},
archivePrefix = {arXiv},
       eprint = {2512.03230},
 primaryClass = {astro-ph.CO},
       adsurl = {https://ui.adsabs.harvard.edu/abs/2025arXiv251203230T}
}

@ARTICLE{2025JCAP...09..053I,
       author = {{Ishak}, M. and {Pan}, J. and {Calderon}, R. and {Lodha}, K. and {Valogiannis}, G. and {Aviles}, A. and {Niz}, G. and {Yi}, L. and {Zheng}, C. and {Garcia-Quintero}, C. and et al.},
        title = "{Modified gravity constraints from the full shape modeling of clustering measurements from DESI 2024}",
      journal = {\jcap},
         year = 2025,
        month = sep,
       volume = {2025},
       number = {9},
          eid = {053},
        pages = {053},
          doi = {10.1088/1475-7516/2025/09/053},
archivePrefix = {arXiv},
       eprint = {2411.12026},
 primaryClass = {astro-ph.CO},
       adsurl = {https://ui.adsabs.harvard.edu/abs/2025JCAP...09..053I}
}

@ARTICLE{2017ApJ...848L..13A,
       author = {{Abbott}, B.~P. and {Abbott}, R. and {Abbott}, T.~D. and {Acernese}, F. and {Ackley}, K. and {Adams}, C. and {Adams}, T. and {Addesso}, P. and {Adhikari}, R.~X. and {Adya}, V.~B. and et al.},
        title = "{Gravitational Waves and Gamma-Rays from a Binary Neutron Star Merger: GW170817 and GRB 170817A}",
      journal = {\apjl},
         year = 2017,
        month = oct,
       volume = {848},
       number = {2},
          eid = {L13},
        pages = {L13},
          doi = {10.3847/2041-8213/aa920c},
archivePrefix = {arXiv},
       eprint = {1710.05834},
 primaryClass = {astro-ph.HE},
       adsurl = {https://ui.adsabs.harvard.edu/abs/2017ApJ...848L..13A}
}

@ARTICLE{2021JCAP...05..057T,
       author = {{Torrado}, Jes{\'u}s and {Lewis}, Antony},
        title = "{Cobaya: code for Bayesian analysis of hierarchical physical models}",
      journal = {\jcap},
         year = 2021,
        month = may,
       volume = {2021},
       number = {5},
          eid = {057},
        pages = {057},
          doi = {10.1088/1475-7516/2021/05/057},
archivePrefix = {arXiv},
       eprint = {2005.05290},
 primaryClass = {astro-ph.IM},
       adsurl = {https://ui.adsabs.harvard.edu/abs/2021JCAP...05..057T}
}

@ARTICLE{1992StaSc...7..457G,
       author = {{Gelman}, Andrew and {Rubin}, Donald B.},
        title = "{Inference from Iterative Simulation Using Multiple Sequences}",
      journal = {Statistical Science},
         year = 1992,
        month = jan,
       volume = {7},
        pages = {457-472},
          doi = {10.1214/ss/1177011136},
       adsurl = {https://ui.adsabs.harvard.edu/abs/1992StaSc...7..457G}
}

@ARTICLE{2013PhRvD..87j3529L,
       author = {{Lewis}, Antony},
        title = "{Efficient sampling of fast and slow cosmological parameters}",
      journal = {\prd},
         year = 2013,
        month = may,
       volume = {87},
       number = {10},
          eid = {103529},
        pages = {103529},
          doi = {10.1103/PhysRevD.87.103529},
archivePrefix = {arXiv},
       eprint = {1304.4473},
 primaryClass = {astro-ph.CO},
       adsurl = {https://ui.adsabs.harvard.edu/abs/2013PhRvD..87j3529L}
}

@ARTICLE{2017JCAP...08..019Z,
       author = {{Zumalac{\'a}rregui}, Miguel and {Bellini}, Emilio and {Sawicki}, Ignacy and {Lesgourgues}, Julien and {Ferreira}, Pedro G.},
        title = "{hi\_class: Horndeski in the Cosmic Linear Anisotropy Solving System}",
      journal = {\jcap},
         year = 2017,
        month = aug,
       volume = {2017},
       number = {8},
          eid = {019},
        pages = {019},
          doi = {10.1088/1475-7516/2017/08/019},
archivePrefix = {arXiv},
       eprint = {1605.06102},
 primaryClass = {astro-ph.CO},
       adsurl = {https://ui.adsabs.harvard.edu/abs/2017JCAP...08..019Z}
}

@ARTICLE{2019PhRvD..99j3502N,
       author = {{Noller}, Johannes and {Nicola}, Andrina},
        title = "{Cosmological parameter constraints for Horndeski scalar-tensor gravity}",
      journal = {\prd},
         year = 2019,
        month = may,
       volume = {99},
       number = {10},
          eid = {103502},
        pages = {103502},
          doi = {10.1103/PhysRevD.99.103502},
archivePrefix = {arXiv},
       eprint = {1811.12928},
 primaryClass = {astro-ph.CO},
       adsurl = {https://ui.adsabs.harvard.edu/abs/2019PhRvD..99j3502N}
}

@ARTICLE{2023MNRAS.520..161X,
       author = {{Xu}, Haojie and {Zhang}, Pengjie and {Peng}, Hui and {Yu}, Yu and {Zhang}, Le and {Yao}, Ji and {Qin}, Jian and {Sun}, Zeyang and {He}, Min and {Yang}, Xiaohu},
        title = "{Using angular two-point correlations to self-calibrate the photometric redshift distributions of DECaLS DR9}",
      journal = {\mnras},
         year = 2023,
        month = mar,
       volume = {520},
       number = {1},
        pages = {161-179},
          doi = {10.1093/mnras/stad136},
archivePrefix = {arXiv},
       eprint = {2209.03967},
 primaryClass = {astro-ph.CO},
       adsurl = {https://ui.adsabs.harvard.edu/abs/2023MNRAS.520..161X}
}

@ARTICLE{2021MNRAS.501.3309Z,
       author = {{Zhou}, Rongpu and {Newman}, Jeffrey A. and {Mao}, Yao-Yuan and {Meisner}, Aaron and {Moustakas}, John and {Myers}, Adam D. and {Prakash}, Abhishek and {Zentner}, Andrew R. and {Brooks}, David and {Duan}, Yutong and et al.},
        title = "{The clustering of DESI-like luminous red galaxies using photometric redshifts}",
      journal = {\mnras},
         year = 2021,
        month = mar,
       volume = {501},
       number = {3},
        pages = {3309-3331},
          doi = {10.1093/mnras/staa3764},
archivePrefix = {arXiv},
       eprint = {2001.06018},
 primaryClass = {astro-ph.CO},
       adsurl = {https://ui.adsabs.harvard.edu/abs/2021MNRAS.501.3309Z}
}

@ARTICLE{1967ApJ...147...73S,
       author = {{Sachs}, R.~K. and {Wolfe}, A.~M.},
        title = "{Perturbations of a Cosmological Model and Angular Variations of the Microwave Background}",
      journal = {\apj},
         year = 1967,
        month = jan,
       volume = {147},
        pages = {73},
          doi = {10.1086/148982},
       adsurl = {https://ui.adsabs.harvard.edu/abs/1967ApJ...147...73S}
}

@ARTICLE{2021MNRAS.500.3838D,
       author = {{Dong}, Fuyu and {Yu}, Yu and {Zhang}, Jun and {Yang}, Xiaohu and {Zhang}, Pengjie},
        title = "{Measuring the integrated Sachs-Wolfe effect from the low-density regions of the universe}",
      journal = {\mnras},
         year = 2021,
        month = jan,
       volume = {500},
       number = {3},
        pages = {3838-3853},
          doi = {10.1093/mnras/staa3194},
archivePrefix = {arXiv},
       eprint = {2006.14202},
 primaryClass = {astro-ph.CO},
       adsurl = {https://ui.adsabs.harvard.edu/abs/2021MNRAS.500.3838D}
}

@ARTICLE{2022MNRAS.517.3785B,
       author = {{Bahr-Kalus}, Benedict and {Parkinson}, David and {Asorey}, Jacobo and {Camera}, Stefano and {Hale}, Catherine and {Qin}, Fei},
        title = "{A measurement of the integrated Sachs-Wolfe effect with the Rapid ASKAP Continuum Survey}",
      journal = {\mnras},
         year = 2022,
        month = dec,
       volume = {517},
       number = {3},
        pages = {3785-3803},
          doi = {10.1093/mnras/stac2040},
archivePrefix = {arXiv},
       eprint = {2204.13436},
 primaryClass = {astro-ph.CO},
       adsurl = {https://ui.adsabs.harvard.edu/abs/2022MNRAS.517.3785B}
}

@ARTICLE{2023ApJ...958..180A,
       author = {{Appleby}, Stephen and {Tonegawa}, Motonari and {Park}, Changbom and {Hong}, Sungwook E. and {Kim}, Juhan and {Yoon}, Yongmin},
        title = "{Cosmological Parameter Constraints from the SDSS Density and Momentum Power Spectra}",
      journal = {\apj},
         year = 2023,
        month = dec,
       volume = {958},
       number = {2},
          eid = {180},
        pages = {180},
          doi = {10.3847/1538-4357/acff68},
archivePrefix = {arXiv},
       eprint = {2305.01943},
 primaryClass = {astro-ph.CO},
       adsurl = {https://ui.adsabs.harvard.edu/abs/2023ApJ...958..180A}
}

@ARTICLE{2017MNRAS.471.3135H,
       author = {{Howlett}, Cullan and {Staveley-Smith}, Lister and {Elahi}, Pascal J. and {Hong}, Tao and {Jarrett}, Tom H. and {Jones}, D. Heath and {Koribalski}, B{\"a}rbel S. and {Macri}, Lucas M. and {Masters}, Karen L. and {Springob}, Christopher M.},
        title = "{2MTF - VI. Measuring the velocity power spectrum}",
      journal = {\mnras},
         year = 2017,
        month = nov,
       volume = {471},
       number = {3},
        pages = {3135-3151},
          doi = {10.1093/mnras/stx1521},
archivePrefix = {arXiv},
       eprint = {1706.05130},
 primaryClass = {astro-ph.CO},
       adsurl = {https://ui.adsabs.harvard.edu/abs/2017MNRAS.471.3135H}
}

@ARTICLE{2019MNRAS.487.5235Q,
       author = {{Qin}, Fei and {Howlett}, Cullan and {Staveley-Smith}, Lister},
        title = "{The redshift-space momentum power spectrum - II. Measuring the growth rate from the combined 2MTF and 6dFGSv surveys}",
      journal = {\mnras},
         year = 2019,
        month = aug,
       volume = {487},
       number = {4},
        pages = {5235-5247},
          doi = {10.1093/mnras/stz1576},
archivePrefix = {arXiv},
       eprint = {1906.02874},
 primaryClass = {astro-ph.CO},
       adsurl = {https://ui.adsabs.harvard.edu/abs/2019MNRAS.487.5235Q}
}

@ARTICLE{2020MNRAS.497.1275S,
       author = {{Said}, Khaled and {Colless}, Matthew and {Magoulas}, Christina and {Lucey}, John R. and {Hudson}, Michael J.},
        title = "{Joint analysis of 6dFGS and SDSS peculiar velocities for the growth rate of cosmic structure and tests of gravity}",
      journal = {\mnras},
         year = 2020,
        month = sep,
       volume = {497},
       number = {1},
        pages = {1275-1293},
          doi = {10.1093/mnras/staa2032},
archivePrefix = {arXiv},
       eprint = {2007.04993},
 primaryClass = {astro-ph.CO},
       adsurl = {https://ui.adsabs.harvard.edu/abs/2020MNRAS.497.1275S}
}

@ARTICLE{2021PhRvD.103h3533A,
       author = {{Alam}, Shadab and {Aubert}, Marie and {Avila}, Santiago and {Balland}, Christophe and {Bautista}, Julian E. and {Bershady}, Matthew A. and {Bizyaev}, Dmitry and {Blanton}, Michael R. and {Bolton}, Adam S. and {Bovy}, Jo and et al.},
        title = "{Completed SDSS-IV extended Baryon Oscillation Spectroscopic Survey: Cosmological implications from two decades of spectroscopic surveys at the Apache Point Observatory}",
      journal = {\prd},
         year = 2021,
        month = apr,
       volume = {103},
       number = {8},
          eid = {083533},
        pages = {083533},
          doi = {10.1103/PhysRevD.103.083533},
archivePrefix = {arXiv},
       eprint = {2007.08991},
 primaryClass = {astro-ph.CO},
       adsurl = {https://ui.adsabs.harvard.edu/abs/2021PhRvD.103h3533A}
}

@ARTICLE{2024MNRAS.531...84B,
       author = {{Boubel}, Paula and {Colless}, Matthew and {Said}, Khaled and {Staveley-Smith}, Lister},
        title = "{Large-scale motions and growth rate from forward-modelling Tully-Fisher peculiar velocities}",
      journal = {\mnras},
         year = 2024,
        month = jun,
       volume = {531},
       number = {1},
        pages = {84-109},
          doi = {10.1093/mnras/stae1122},
archivePrefix = {arXiv},
       eprint = {2301.12648},
 primaryClass = {astro-ph.CO},
       adsurl = {https://ui.adsabs.harvard.edu/abs/2024MNRAS.531...84B}
}

@ARTICLE{2020A&A...641A...1P,
       author = {{Planck Collaboration} and {Aghanim}, N. and {Akrami}, Y. and {Arroja}, F. and {Ashdown}, M. and {Aumont}, J. and {Baccigalupi}, C. and {Ballardini}, M. and {Banday}, A.~J. and {Barreiro}, R.~B. and et al.},
        title = "{Planck 2018 results. I. Overview and the cosmological legacy of Planck}",
      journal = {\aap},
         year = 2020,
        month = sep,
       volume = {641},
          eid = {A1},
        pages = {A1},
          doi = {10.1051/0004-6361/201833880},
archivePrefix = {arXiv},
       eprint = {1807.06205},
 primaryClass = {astro-ph.CO},
       adsurl = {https://ui.adsabs.harvard.edu/abs/2020A&A...641A...1P}
}

@ARTICLE{1998AJ....116.1009R,
       author = {{Riess}, Adam G. and {Filippenko}, Alexei V. and {Challis}, Peter and {Clocchiatti}, Alejandro and {Diercks}, Alan and {Garnavich}, Peter M. and {Gilliland}, Ron L. and {Hogan}, Craig J. and {Jha}, Saurabh and {Kirshner}, Robert P. and et al.},
        title = "{Observational Evidence from Supernovae for an Accelerating Universe and a Cosmological Constant}",
      journal = {\aj},
         year = 1998,
        month = sep,
       volume = {116},
       number = {3},
        pages = {1009-1038},
          doi = {10.1086/300499},
archivePrefix = {arXiv},
       eprint = {astro-ph/9805201},
 primaryClass = {astro-ph},
       adsurl = {https://ui.adsabs.harvard.edu/abs/1998AJ....116.1009R}
}

@ARTICLE{2020A&A...641A...6P,
       author = {{Planck Collaboration} and {Aghanim}, N. and {Akrami}, Y. and {Ashdown}, M. and {Aumont}, J. and {Baccigalupi}, C. and {Ballardini}, M. and {Banday}, A.~J. and {Barreiro}, R.~B. and {Bartolo}, N. and et al.},
        title = "{Planck 2018 results. VI. Cosmological parameters}",
      journal = {\aap},
         year = 2020,
        month = sep,
       volume = {641},
          eid = {A6},
        pages = {A6},
          doi = {10.1051/0004-6361/201833910},
archivePrefix = {arXiv},
       eprint = {1807.06209},
 primaryClass = {astro-ph.CO},
       adsurl = {https://ui.adsabs.harvard.edu/abs/2020A&A...641A...6P}
}

@ARTICLE{2012PhR...513....1C,
       author = {{Clifton}, Timothy and {Ferreira}, Pedro G. and {Padilla}, Antonio and {Skordis}, Constantinos},
        title = "{Modified gravity and cosmology}",
      journal = {\physrep},
         year = 2012,
        month = mar,
       volume = {513},
       number = {1},
        pages = {1-189},
          doi = {10.1016/j.physrep.2012.01.001},
archivePrefix = {arXiv},
       eprint = {1106.2476},
 primaryClass = {astro-ph.CO},
       adsurl = {https://ui.adsabs.harvard.edu/abs/2012PhR...513....1C}
}

@ARTICLE{2016RPPh...79d6902K,
       author = {{Koyama}, Kazuya},
        title = "{Cosmological tests of modified gravity}",
      journal = {Reports on Progress in Physics},
         year = 2016,
        month = apr,
       volume = {79},
       number = {4},
          eid = {046902},
        pages = {046902},
          doi = {10.1088/0034-4885/79/4/046902},
archivePrefix = {arXiv},
       eprint = {1504.04623},
 primaryClass = {astro-ph.CO},
       adsurl = {https://ui.adsabs.harvard.edu/abs/2016RPPh...79d6902K}
}

@ARTICLE{2019LRR....22....1I,
       author = {{Ishak}, Mustapha},
        title = "{Testing general relativity in cosmology}",
      journal = {Living Reviews in Relativity},
         year = 2019,
        month = dec,
       volume = {22},
       number = {1},
          eid = {1},
        pages = {1},
          doi = {10.1007/s41114-018-0017-4},
archivePrefix = {arXiv},
       eprint = {1806.10122},
 primaryClass = {astro-ph.CO},
       adsurl = {https://ui.adsabs.harvard.edu/abs/2019LRR....22....1I}
}

@ARTICLE{2019ARA&A..57..335F,
       author = {{Ferreira}, Pedro G.},
        title = "{Cosmological Tests of Gravity}",
      journal = {\araa},
         year = 2019,
        month = aug,
       volume = {57},
        pages = {335-374},
          doi = {10.1146/annurev-astro-091918-104423},
archivePrefix = {arXiv},
       eprint = {1902.10503},
 primaryClass = {astro-ph.CO},
       adsurl = {https://ui.adsabs.harvard.edu/abs/2019ARA&A..57..335F}
}

@ARTICLE{2026arXiv260210065D,
       author = {{DES Collaboration} and {Abbott}, T.~M.~C. and {Aguena}, M. and {Alarcon}, A. and {Alves}, O. and {Amon}, A. and {Anbajagane}, D. and {Andrade-Oliveira}, F. and {d'Assignies}, W. and {Avila}, S. and et al.},
        title = "{Dark Energy Survey Year 6 Results: Cosmological Constraints from Cosmic Shear}",
      journal = {arXiv e-prints},
         year = 2026,
        month = feb,
          eid = {arXiv:2602.10065},
        pages = {arXiv:2602.10065},
          doi = {10.48550/arXiv.2602.10065},
archivePrefix = {arXiv},
       eprint = {2602.10065},
 primaryClass = {astro-ph.CO},
       adsurl = {https://ui.adsabs.harvard.edu/abs/2026arXiv260210065D}
}

@ARTICLE{2026arXiv260114559D,
       author = {{DES Collaboration} and {Abbott}, T.~M.~C. and {Adamow}, M. and {Aguena}, M. and {Alarcon}, A. and {Allam}, S.~S. and {Alves}, O. and {Amon}, A. and {Anbajagane}, D. and {Andrade-Oliveira}, F. and et al.},
        title = "{Dark Energy Survey Year 6 Results: Cosmological Constraints from Galaxy Clustering and Weak Lensing}",
      journal = {arXiv e-prints},
         year = 2026,
        month = jan,
          eid = {arXiv:2601.14559},
        pages = {arXiv:2601.14559},
          doi = {10.48550/arXiv.2601.14559},
archivePrefix = {arXiv},
       eprint = {2601.14559},
 primaryClass = {astro-ph.CO},
       adsurl = {https://ui.adsabs.harvard.edu/abs/2026arXiv260114559D}
}

@ARTICLE{2022ApJ...938..110B,
       author = {{Brout}, Dillon and {Scolnic}, Dan and {Popovic}, Brodie and {Riess}, Adam G. and {Carr}, Anthony and {Zuntz}, Joe and {Kessler}, Rick and {Davis}, Tamara M. and {Hinton}, Samuel and {Jones}, David and et al.},
        title = "{The Pantheon+ Analysis: Cosmological Constraints}",
      journal = {\apj},
         year = 2022,
        month = oct,
       volume = {938},
       number = {2},
          eid = {110},
        pages = {110},
          doi = {10.3847/1538-4357/ac8e04},
archivePrefix = {arXiv},
       eprint = {2202.04077},
 primaryClass = {astro-ph.CO},
       adsurl = {https://ui.adsabs.harvard.edu/abs/2022ApJ...938..110B}
}

@ARTICLE{2025JCAP...02..021A,
       author = {{Adame}, A.~G. and {Aguilar}, J. and {Ahlen}, S. and {Alam}, S. and {Alexander}, D.~M. and {Alvarez}, M. and {Alves}, O. and {Anand}, A. and {Andrade}, U. and {Armengaud}, E. and et al.},
        title = "{DESI 2024 VI: cosmological constraints from the measurements of baryon acoustic oscillations}",
      journal = {\jcap},
         year = 2025,
        month = feb,
       volume = {2025},
       number = {2},
          eid = {021},
        pages = {021},
          doi = {10.1088/1475-7516/2025/02/021},
archivePrefix = {arXiv},
       eprint = {2404.03002},
 primaryClass = {astro-ph.CO},
       adsurl = {https://ui.adsabs.harvard.edu/abs/2025JCAP...02..021A}
}

@ARTICLE{2025A&A...702A.169S,
       author = {{St{\"o}lzner}, Benjamin and {Wright}, Angus H. and {Asgari}, Marika and {Heymans}, Catherine and {Hildebrandt}, Hendrik and {Hoekstra}, Henk and {Joachimi}, Benjamin and {Kuijken}, Konrad and {Li}, Shun-Sheng and {Mahony}, Constance and et al.},
        title = "{KiDS-Legacy: Consistency of cosmic shear measurements and joint cosmological constraints with external probes}",
      journal = {\aap},
         year = 2025,
        month = oct,
       volume = {702},
          eid = {A169},
        pages = {A169},
          doi = {10.1051/0004-6361/202554893},
archivePrefix = {arXiv},
       eprint = {2503.19442},
 primaryClass = {astro-ph.CO},
       adsurl = {https://ui.adsabs.harvard.edu/abs/2025A&A...702A.169S}
}

@ARTICLE{2025PhRvD.112h3515A,
       author = {{Abdul Karim}, M. and {Aguilar}, J. and {Ahlen}, S. and {Alam}, S. and {Allen}, L. and {Allende Prieto}, C. and {Alves}, O. and {Anand}, A. and {Andrade}, U. and {Armengaud}, E. and et al.},
        title = "{DESI DR2 results. II. Measurements of baryon acoustic oscillations and cosmological constraints}",
      journal = {\prd},
         year = 2025,
        month = oct,
       volume = {112},
       number = {8},
          eid = {083515},
        pages = {083515},
          doi = {10.1103/tr6y-kpc6},
archivePrefix = {arXiv},
       eprint = {2503.14738},
 primaryClass = {astro-ph.CO},
       adsurl = {https://ui.adsabs.harvard.edu/abs/2025PhRvD.112h3515A}
}

@ARTICLE{2015PhR...568....1J,
       author = {{Joyce}, Austin and {Jain}, Bhuvnesh and {Khoury}, Justin and {Trodden}, Mark},
        title = "{Beyond the cosmological standard model}",
      journal = {\physrep},
         year = 2015,
        month = mar,
       volume = {568},
        pages = {1-98},
          doi = {10.1016/j.physrep.2014.12.002},
archivePrefix = {arXiv},
       eprint = {1407.0059},
 primaryClass = {astro-ph.CO},
       adsurl = {https://ui.adsabs.harvard.edu/abs/2015PhR...568....1J}
}

@ARTICLE{2016A&A...594A..21P,
       author = {{Planck Collaboration} and {Ade}, P.~A.~R. and {Aghanim}, N. and {Arnaud}, M. and {Ashdown}, M. and {Aumont}, J. and {Baccigalupi}, C. and {Banday}, A.~J. and {Barreiro}, R.~B. and {Bartolo}, N. and et al.},
        title = "{Planck 2015 results. XXI. The integrated Sachs-Wolfe effect}",
      journal = {\aap},
         year = 2016,
        month = sep,
       volume = {594},
          eid = {A21},
        pages = {A21},
          doi = {10.1051/0004-6361/201525831},
archivePrefix = {arXiv},
       eprint = {1502.01595},
 primaryClass = {astro-ph.CO},
       adsurl = {https://ui.adsabs.harvard.edu/abs/2016A&A...594A..21P}
}

@ARTICLE{2025PhRvD.112h3537C,
       author = {{Chudaykin}, Anton and {Kunz}, Martin and {Carron}, Julien},
        title = "{Modified gravity constraints with the Planck ISW-lensing bispectrum}",
      journal = {\prd},
         year = 2025,
        month = oct,
       volume = {112},
       number = {8},
          eid = {083537},
        pages = {083537},
          doi = {10.1103/9zjh-8htv},
archivePrefix = {arXiv},
       eprint = {2503.09893},
 primaryClass = {astro-ph.CO},
       adsurl = {https://ui.adsabs.harvard.edu/abs/2025PhRvD.112h3537C}
}

@ARTICLE{2024PhRvD.110l3525S,
       author = {{Seraille}, Emeric and {Noller}, Johannes and {Sherwin}, Blake D.},
        title = "{Constraining dark energy with the integrated Sachs-Wolfe effect}",
      journal = {\prd},
         year = 2024,
        month = dec,
       volume = {110},
       number = {12},
          eid = {123525},
        pages = {123525},
          doi = {10.1103/PhysRevD.110.123525},
archivePrefix = {arXiv},
       eprint = {2401.06221},
 primaryClass = {astro-ph.CO},
       adsurl = {https://ui.adsabs.harvard.edu/abs/2024PhRvD.110l3525S}
}

@ARTICLE{2022JCAP...09..002K,
       author = {{Kable}, Joshua A. and {Benevento}, Giampaolo and {Frusciante}, Noemi and {De Felice}, Antonio and {Tsujikawa}, Shinji},
        title = "{Probing modified gravity with integrated Sachs-Wolfe CMB and galaxy cross-correlations}",
      journal = {\jcap},
         year = 2022,
        month = sep,
       volume = {2022},
       number = {9},
          eid = {002},
        pages = {002},
          doi = {10.1088/1475-7516/2022/09/002},
archivePrefix = {arXiv},
       eprint = {2111.10432},
 primaryClass = {astro-ph.CO},
       adsurl = {https://ui.adsabs.harvard.edu/abs/2022JCAP...09..002K}
}

@ARTICLE{2023OJAp....6E..36D,
       author = {{Dark Energy Survey and Kilo-Degree Survey Collaboration} and {Abbott}, T.~M.~C. and {Aguena}, M. and {Alarcon}, A. and {Alves}, O. and {Amon}, A. and {Andrade-Oliveira}, F. and {Asgari}, M. and {Avila}, S. and {Bacon}, D. and {Bechtol}, K. and {Becker}, M.~R. and {Bernstein}, G.~M. and {Bertin}, E. and {Bilicki}, M. and {Blazek}, J. and {Bocquet}, S. and {Brooks}, D. and {Burger}, P. and {Burke}, D.~L. and {Camacho}, H. and {Campos}, A. and {Carnero Rosell}, A. and {Carrasco Kind}, M. and {Carretero}, J. and {Castander}, F.~J. and {Cawthon}, R. and {Chang}, C. and {Chen}, R. and {Choi}, A. and {Conselice}, C. and {Cordero}, J. and {Crocce}, M. and {da Costa}, L.~N. and {da Silva Pereira}, M.~E. and {Dalal}, R. and {Davis}, C. and {de Jong}, J.~T.~A. and {DeRose}, J. and {Desai}, S. and {Diehl}, H.~T. and {Dodelson}, S. and {Doel}, P. and {Doux}, C. and {Drlica-Wagner}, A. and {Dvornik}, A. and {Eckert}, K. and {Eifler}, T.~F. and {Elvin-Poole}, J. and {Everett}, S. and {Fang}, X. and {Ferrero}, I. and {Fert{\'e}}, A. and {Flaugher}, B. and {Friedrich}, O. and {Frieman}, J. and {Garc{\'\i}a-Bellido}, J. and {Gatti}, M. and {Giannini}, G. and {Giblin}, B. and {Gruen}, D. and {Gruendl}, R.~A. and {Gutierrez}, G. and {Harrison}, I. and {Hartley}, W.~G. and {Herner}, K. and {Heymans}, C. and {Hildebrandt}, H. and {Hinton}, S.~R. and {Hoekstra}, H. and {Hollowood}, D.~L. and {Honscheid}, K. and {Huang}, H. and {Huff}, E.~M. and {Huterer}, D. and {James}, D.~J. and {Jarvis}, M. and {Jeffrey}, N. and {Jeltema}, T. and {Joachimi}, B. and {Joudaki}, S. and {Kannawadi}, A. and {Krause}, E. and {Kuehn}, K. and {Kuijken}, K. and {Kuropatkin}, N. and {Lahav}, O. and {Leget}, P.-F. and {Lemos}, P. and {Li}, S.-S. and {Li}, X. and {Liddle}, A.~R. and {Lima}, M. and {Lin}, C.-A. and {Lin}, H. and {MacCrann}, N. and {Mahony}, C. and {Marshall}, J.~L. and {McCullough}, J. and {Mena-Fern{\'a}ndez}, J. and {Menanteau}, F. and {Miquel}, R. and {Mohr}, J.~J. and {Muir}, J. and {Myles}, J. and {Napolitano}, N. and {Navarro-Alsina}, A. and {Ogando}, R.~L.~C. and {Palmese}, A. and {Pandey}, S. and {Park}, Y. and {Paterno}, M. and {Peacock}, J.~A. and {Petravick}, D. and {Pieres}, A. and {Plazas Malag{\'o}n}, A.~A. and {Porredon}, A. and {Prat}, J. and {Radovich}, M. and {Raveri}, M. and {Reischke}, R. and {Robertson}, N.~C. and {Rollins}, R.~P. and {Romer}, A.~K. and {Roodman}, A. and {Rykoff}, E.~S. and {Samuroff}, S. and {S{\'a}nchez}, C. and {Sanchez}, E. and {Sanchez}, J. and {Schneider}, P. and {Secco}, L.~F. and {Sevilla-Noarbe}, I. and {Shan}, H.-Y. and {Sheldon}, E. and {Shin}, T. and {Sif{\'o}n}, C. and {Smith}, M. and {Soares-Santos}, M. and {St{\"o}lzner}, B. and {Suchyta}, E. and {Swanson}, M.~E.~C. and {Tarle}, G. and {Thomas}, D. and {To}, C. and {Troxel}, M.~A. and {Tr{\"o}ster}, T. and {Tutusaus}, I. and {van den Busch}, J.~L. and {Varga}, T.~N. and {Walker}, A.~R. and {Weaverdyck}, N. and {Wechsler}, R.~H. and {Weller}, J. and {Wiseman}, P. and {Wright}, A.~H. and {Yanny}, B. and {Yin}, B. and {Yoon}, M. and {Zhang}, Y. and {Zuntz}, J.},
        title = "{DES Y3 + KiDS-1000: Consistent cosmology combining cosmic shear surveys}",
      journal = {The Open Journal of Astrophysics},
         year = 2023,
        month = oct,
       volume = {6},
          eid = {36},
        pages = {36},
          doi = {10.21105/astro.2305.17173},
archivePrefix = {arXiv},
       eprint = {2305.17173},
 primaryClass = {astro-ph.CO},
       adsurl = {https://ui.adsabs.harvard.edu/abs/2023OJAp....6E..36D}
}

@ARTICLE{2023MNRAS.520.5016L,
       author = {{Longley}, Emily P. and {Chang}, Chihway and {Walter}, Christopher W. and {Zuntz}, Joe and {Ishak}, Mustapha and {Mandelbaum}, Rachel and {Miyatake}, Hironao and {Nicola}, Andrina and {Pedersen}, Eske M. and {Pereira}, Maria E.~S. and {Prat}, Judit and {S{\'a}nchez}, J. and {Secco}, Lucas F. and {Tr{\"o}ster}, Tilman and {Troxel}, Michael and {Wright}, Angus H. and {LSST Dark Energy Science Collaboration}},
        title = "{A unified catalogue-level reanalysis of stage-III cosmic shear surveys}",
      journal = {\mnras},
         year = 2023,
        month = apr,
       volume = {520},
       number = {4},
        pages = {5016-5041},
          doi = {10.1093/mnras/stad246},
archivePrefix = {arXiv},
       eprint = {2208.07179},
 primaryClass = {astro-ph.CO},
       adsurl = {https://ui.adsabs.harvard.edu/abs/2023MNRAS.520.5016L}
}

@ARTICLE{2019MNRAS.489..401Z,
       author = {{Zubeldia}, {\'I}{\~n}igo and {Challinor}, Anthony},
        title = "{Cosmological constraints from Planck galaxy clusters with CMB lensing mass bias calibration}",
      journal = {\mnras},
         year = 2019,
        month = oct,
       volume = {489},
       number = {1},
        pages = {401-419},
          doi = {10.1093/mnras/stz2153},
archivePrefix = {arXiv},
       eprint = {1904.07887},
 primaryClass = {astro-ph.CO},
       adsurl = {https://ui.adsabs.harvard.edu/abs/2019MNRAS.489..401Z}
}

@ARTICLE{2024PhRvD.110h3510B,
       author = {{Bocquet}, S. and {Grandis}, S. and {Bleem}, L.~E. and {Klein}, M. and {Mohr}, J.~J. and {Schrabback}, T. and {Abbott}, T.~M.~C. and {Ade}, P.~A.~R. and {Aguena}, M. and {Alarcon}, A. and {Allam}, S. and {Allen}, S.~W. and {Alves}, O. and {Amon}, A. and {Anderson}, A.~J. and {Annis}, J. and {Ansarinejad}, B. and {Austermann}, J.~E. and {Avila}, S. and {Bacon}, D. and {Bayliss}, M. and {Beall}, J.~A. and {Bechtol}, K. and {Becker}, M.~R. and {Bender}, A.~N. and {Benson}, B.~A. and {Bernstein}, G.~M. and {Bhargava}, S. and {Bianchini}, F. and {Brodwin}, M. and {Brooks}, D. and {Bryant}, L. and {Campos}, A. and {Canning}, R.~E.~A. and {Carlstrom}, J.~E. and {Carnero Rosell}, A. and {Carrasco Kind}, M. and {Carretero}, J. and {Castander}, F.~J. and {Cawthon}, R. and {Chang}, C.~L. and {Chang}, C. and {Chaubal}, P. and {Chen}, R. and {Chiang}, H.~C. and {Choi}, A. and {Chou}, T.-L. and {Citron}, R. and {Corbett Moran}, C. and {Cordero}, J. and {Costanzi}, M. and {Crawford}, T.~M. and {Crites}, A.~T. and {da Costa}, L.~N. and {Pereira}, M.~E.~S. and {Davis}, C. and {Davis}, T.~M. and {DeRose}, J. and {Desai}, S. and {de Haan}, T. and {Diehl}, H.~T. and {Dobbs}, M.~A. and {Dodelson}, S. and {Doux}, C. and {Drlica-Wagner}, A. and {Eckert}, K. and {Elvin-Poole}, J. and {Everett}, S. and {Everett}, W. and {Ferrero}, I. and {Fert{\'e}}, A. and {Flores}, A.~M. and {Frieman}, J. and {Gallicchio}, J. and {Garc{\'\i}a-Bellido}, J. and {Gatti}, M. and {George}, E.~M. and {Giannini}, G. and {Gladders}, M.~D. and {Gruen}, D. and {Gruendl}, R.~A. and {Gupta}, N. and {Gutierrez}, G. and {Halverson}, N.~W. and {Harrison}, I. and {Hartley}, W.~G. and {Herner}, K. and {Hinton}, S.~R. and {Holder}, G.~P. and {Hollowood}, D.~L. and {Holzapfel}, W.~L. and {Honscheid}, K. and {Hrubes}, J.~D. and {Huang}, N. and {Hubmayr}, J. and {Huff}, E.~M. and {Huterer}, D. and {Irwin}, K.~D. and {James}, D.~J. and {Jarvis}, M. and {Khullar}, G. and {Kim}, K. and {Knox}, L. and {Kraft}, R. and {Krause}, E. and {Kuehn}, K. and {Kuropatkin}, N. and {K{\'e}ruzor{\'e}}, F. and {Lahav}, O. and {Lee}, A.~T. and {Leget}, P.-F. and {Li}, D. and {Lin}, H. and {Lowitz}, A. and {MacCrann}, N. and {Mahler}, G. and {Mantz}, A. and {Marshall}, J.~L. and {McCullough}, J. and {McDonald}, M. and {McMahon}, J.~J. and {Mena-Fern{\'a}ndez}, J. and {Menanteau}, F. and {Meyer}, S.~S. and {Miquel}, R. and {Montgomery}, J. and {Myles}, J. and {Natoli}, T. and {Navarro-Alsina}, A. and {Nibarger}, J.~P. and {Noble}, G.~I. and {Novosad}, V. and {Ogando}, R.~L.~C. and {Omori}, Y. and {Padin}, S. and {Pandey}, S. and {Paschos}, P. and {Patil}, S. and {Pieres}, A. and {Plazas Malag{\'o}n}, A.~A. and {Porredon}, A. and {Prat}, J. and {Pryke}, C. and {Raveri}, M. and {Reichardt}, C.~L. and {Roberson}, J. and {Rollins}, R.~P. and {Romero}, C. and {Roodman}, A. and {Ruhl}, J.~E. and {Rykoff}, E.~S. and {Saliwanchik}, B.~R. and {Salvati}, L. and {S{\'a}nchez}, C. and {Sanchez}, E. and {Sanchez Cid}, D. and {Saro}, A. and {Schaffer}, K.~K. and {Secco}, L.~F. and {Sevilla-Noarbe}, I. and {Sharon}, K. and {Sheldon}, E. and {Shin}, T. and {Sievers}, C. and {Smecher}, G. and {Smith}, M. and {Somboonpanyakul}, T. and {Sommer}, M. and {Stalder}, B. and {Stark}, A.~A. and {Stephen}, J. and {Strazzullo}, V. and {Suchyta}, E. and {Tarle}, G. and {To}, C. and {Troxel}, M.~A. and {Tucker}, C. and {Tutusaus}, I. and {Varga}, T.~N. and {Veach}, T. and {Vieira}, J.~D. and {Vikhlinin}, A. and {von der Linden}, A. and {Wang}, G. and {Weaverdyck}, N. and {Weller}, J. and {Whitehorn}, N. and {Wu}, W.~L.~K. and {Yanny}, B. and {Yefremenko}, V. and {Yin}, B. and {Young}, M. and {Zebrowski}, J.~A. and {Zhang}, Y. and {Zohren}, H. and {Zuntz}, J. and {(SPT} and {DES Collaborations)}},
        title = "{SPT clusters with DES and HST weak lensing. II. Cosmological constraints from the abundance of massive halos}",
      journal = {\prd},
         year = 2024,
        month = oct,
       volume = {110},
       number = {8},
          eid = {083510},
        pages = {083510},
          doi = {10.1103/PhysRevD.110.083510},
archivePrefix = {arXiv},
       eprint = {2401.02075},
 primaryClass = {astro-ph.CO},
       adsurl = {https://ui.adsabs.harvard.edu/abs/2024PhRvD.110h3510B}
}

@ARTICLE{2024PhRvD.110h3511S,
       author = {{Sunayama}, Tomomi and {Miyatake}, Hironao and {Sugiyama}, Sunao and {More}, Surhud and {Li}, Xiangchong and {Dalal}, Roohi and {Rau}, Markus M. and {Shi}, Jingjing and {Chiu}, I.-non and {Shirasaki}, Masato and {Zhang}, Tianqing and {Nishizawa}, Atsushi J.},
        title = "{Optical cluster cosmology with SDSS redMaPPer clusters and HSC-Y3 lensing measurements}",
      journal = {\prd},
         year = 2024,
        month = oct,
       volume = {110},
       number = {8},
          eid = {083511},
        pages = {083511},
          doi = {10.1103/PhysRevD.110.083511},
archivePrefix = {arXiv},
       eprint = {2309.13025},
 primaryClass = {astro-ph.CO},
       adsurl = {https://ui.adsabs.harvard.edu/abs/2024PhRvD.110h3511S}
}

@ARTICLE{2020PhR...857....1F,
       author = {{Frusciante}, Noemi and {Perenon}, Louis},
        title = "{Effective field theory of dark energy: A review}",
      journal = {\physrep},
         year = 2020,
        month = may,
       volume = {857},
        pages = {1-63},
          doi = {10.1016/j.physrep.2020.02.004},
archivePrefix = {arXiv},
       eprint = {1907.03150},
 primaryClass = {astro-ph.CO},
       adsurl = {https://ui.adsabs.harvard.edu/abs/2020PhR...857....1F}
}

@INCOLLECTION{2015LNP...892...97T,
       author = {{Tsujikawa}, Shinji},
        title = "{The Effective Field Theory of Inflation/Dark Energy and the Horndeski Theory}",
    booktitle = {Lecture Notes in Physics, Berlin Springer Verlag},
         year = 2015,
       editor = {{Papantonopoulos}, Eleftherios},
       volume = {892},
        pages = {97},
          doi = {10.1007/978-3-319-10070-8_4},
       adsurl = {https://ui.adsabs.harvard.edu/abs/2015LNP...892...97T}
}

@ARTICLE{2021JCAP...01..013B,
       author = {{Brando}, Guilherme and {Koyama}, Kazuya and {Wands}, David},
        title = "{Relativistic corrections to the growth of structure in modified gravity}",
      journal = {\jcap},
         year = 2021,
        month = jan,
       volume = {2021},
       number = {1},
          eid = {013},
        pages = {013},
          doi = {10.1088/1475-7516/2021/01/013},
archivePrefix = {arXiv},
       eprint = {2006.11019},
 primaryClass = {astro-ph.CO},
       adsurl = {https://ui.adsabs.harvard.edu/abs/2021JCAP...01..013B}
}

@ARTICLE{2021JCAP...09..024B,
       author = {{Brando}, Guilherme and {Koyama}, Kazuya and {Wands}, David and {Zumalac{\'a}rregui}, Miguel and {Sawicki}, Ignacy and {Bellini}, Emilio},
        title = "{Fully relativistic predictions in Horndeski gravity from standard Newtonian N-body simulations}",
      journal = {\jcap},
         year = 2021,
        month = sep,
       volume = {2021},
       number = {9},
          eid = {024},
        pages = {024},
          doi = {10.1088/1475-7516/2021/09/024},
archivePrefix = {arXiv},
       eprint = {2105.04491},
 primaryClass = {astro-ph.CO},
       adsurl = {https://ui.adsabs.harvard.edu/abs/2021JCAP...09..024B}
}

@ARTICLE{2020JCAP...07..015N,
       author = {{Negrelli}, Carolina and {Kraiselburd}, Lucila and {Landau}, Susana and {Sc{\'o}ccola}, Claudia G.},
        title = "{Testing Modified Gravity theory (MOG) with Type Ia Supernovae, Cosmic Chronometers and Baryon Acoustic Oscillations}",
      journal = {\jcap},
         year = 2020,
        month = jul,
       volume = {2020},
       number = {7},
          eid = {015},
        pages = {015},
          doi = {10.1088/1475-7516/2020/07/015},
archivePrefix = {arXiv},
       eprint = {2004.13648},
 primaryClass = {astro-ph.CO},
       adsurl = {https://ui.adsabs.harvard.edu/abs/2020JCAP...07..015N}
}

@ARTICLE{2018MNRAS.480.3725S,
       author = {{Spurio Mancini}, A. and {Reischke}, R. and {Pettorino}, V. and {Sch{\"a}fer}, B.~M. and {Zumalac{\'a}rregui}, M.},
        title = "{Testing (modified) gravity with 3D and tomographic cosmic shear}",
      journal = {\mnras},
         year = 2018,
        month = nov,
       volume = {480},
       number = {3},
        pages = {3725-3738},
          doi = {10.1093/mnras/sty2092},
archivePrefix = {arXiv},
       eprint = {1801.04251},
 primaryClass = {astro-ph.CO},
       adsurl = {https://ui.adsabs.harvard.edu/abs/2018MNRAS.480.3725S}
}

@ARTICLE{2020A&A...642A.158B,
       author = {{Blake}, Chris and {Amon}, Alexandra and {Asgari}, Marika and {Bilicki}, Maciej and {Dvornik}, Andrej and {Erben}, Thomas and {Giblin}, Benjamin and {Glazebrook}, Karl and {Heymans}, Catherine and {Hildebrandt}, Hendrik and et al.},
        title = "{Testing gravity using galaxy-galaxy lensing and clustering amplitudes in KiDS-1000, BOSS, and 2dFLenS}",
      journal = {\aap},
         year = 2020,
        month = oct,
       volume = {642},
          eid = {A158},
        pages = {A158},
          doi = {10.1051/0004-6361/202038505},
archivePrefix = {arXiv},
       eprint = {2005.14351},
 primaryClass = {astro-ph.CO},
       adsurl = {https://ui.adsabs.harvard.edu/abs/2020A&A...642A.158B}
}

@ARTICLE{2017MNRAS.471.1259J,
       author = {{Joudaki}, Shahab and {Mead}, Alexander and {Blake}, Chris and {Choi}, Ami and {de Jong}, Jelte and {Erben}, Thomas and {Fenech Conti}, Ian and {Herbonnet}, Ricardo and {Heymans}, Catherine and {Hildebrandt}, Hendrik and et al.},
        title = "{KiDS-450: testing extensions to the standard cosmological model}",
      journal = {\mnras},
         year = 2017,
        month = oct,
       volume = {471},
       number = {2},
        pages = {1259-1279},
          doi = {10.1093/mnras/stx998},
archivePrefix = {arXiv},
       eprint = {1610.04606},
 primaryClass = {astro-ph.CO},
       adsurl = {https://ui.adsabs.harvard.edu/abs/2017MNRAS.471.1259J}
}

@ARTICLE{2016JCAP...06E.001B,
       author = {{Bellini}, Emilio and {Cuesta}, Antonio J. and {Jimenez}, Raul and {Verde}, Licia},
        title = "{Erratum: Constraints on deviations from {\ensuremath{\Lambda}}CDM within Horndeski gravity Erratum: Constraints on deviations from {\ensuremath{\Lambda}}CDM within Horndeski gravity}",
      journal = {\jcap},
         year = 2016,
        month = jun,
       volume = {2016},
       number = {6},
          eid = {E01},
        pages = {E01},
          doi = {10.1088/1475-7516/2016/06/E01},
       adsurl = {https://ui.adsabs.harvard.edu/abs/2016JCAP...06E.001B}
}

@ARTICLE{2019MNRAS.482.3274R,
       author = {{Reischke}, Robert and {Mancini}, Alessio Spurio and {Sch{\"a}fer}, Bj{\"o}rn Malte and {Merkel}, Philipp M.},
        title = "{Investigating scalar-tensor gravity with statistics of the cosmic large-scale structure}",
      journal = {\mnras},
         year = 2019,
        month = jan,
       volume = {482},
       number = {3},
        pages = {3274-3287},
          doi = {10.1093/mnras/sty2919},
archivePrefix = {arXiv},
       eprint = {1804.02441},
 primaryClass = {astro-ph.CO},
       adsurl = {https://ui.adsabs.harvard.edu/abs/2019MNRAS.482.3274R}
}

@ARTICLE{1999ApJ...517..565P,
       author = {{Perlmutter}, S. and {Aldering}, G. and {Goldhaber}, G. and {Knop}, R.~A. and {Nugent}, P. and {Castro}, P.~G. and {Deustua}, S. and {Fabbro}, S. and {Goobar}, A. and {Groom}, D.~E. and et al.},
        title = "{Measurements of {\ensuremath{\Omega}} and {\ensuremath{\Lambda}} from 42 High-Redshift Supernovae}",
      journal = {\apj},
         year = 1999,
        month = jun,
       volume = {517},
       number = {2},
        pages = {565-586},
          doi = {10.1086/307221},
archivePrefix = {arXiv},
       eprint = {astro-ph/9812133},
 primaryClass = {astro-ph},
       adsurl = {https://ui.adsabs.harvard.edu/abs/1999ApJ...517..565P}
}

@ARTICLE{2025PhRvD.111j3540S,
       author = {{Sailer}, Noah and {DeRose}, Joseph and {Ferraro}, Simone and {Chen}, Shi-Fan and {Zhou}, Rongpu and {White}, Martin and {Kim}, Joshua and {Madhavacheril}, Mathew},
        title = "{Evolution of structure growth during dark energy domination: Insights from the cross-correlation of DESI galaxies with CMB lensing and galaxy magnification}",
      journal = {\prd},
         year = 2025,
        month = may,
       volume = {111},
       number = {10},
          eid = {103540},
        pages = {103540},
          doi = {10.1103/27rg-tq8z},
archivePrefix = {arXiv},
       eprint = {2503.24385},
 primaryClass = {astro-ph.CO},
       adsurl = {https://ui.adsabs.harvard.edu/abs/2025PhRvD.111j3540S}
}

@ARTICLE{2026PhRvD.113f3555L,
       author = {{Lu}, Zhiyu and {Simon}, Th{\'e}o},
        title = "{New multiprobe analysis of modified gravity and evolving dark energy}",
      journal = {\prd},
         year = 2026,
        month = mar,
       volume = {113},
       number = {6},
          eid = {063555},
        pages = {063555},
          doi = {10.1103/hfxk-bdr3},
archivePrefix = {arXiv},
       eprint = {2511.10616},
 primaryClass = {astro-ph.CO},
       adsurl = {https://ui.adsabs.harvard.edu/abs/2026PhRvD.113f3555L}
}

\end{document}